\documentclass[aps,prb,twocolumn,showpacs,floats,epsfig]{revtex4}
\usepackage{epsfig,psfrag,subfigure,amsopn}
\usepackage{graphicx,psfrag,subfigure}
\usepackage{amsmath,amssymb,amsbsy}

\newcommand\beq{\begin{equation}}
\newcommand\eeq{\end{equation}}
\newcommand\bea{\begin{eqnarray}}
\newcommand\eea{\end{eqnarray}}
\newcommand\non{\nonumber}
\newcommand\noi{\noindent}

\newcommand\bib{\bibitem}

\newcommand\al{\alpha}
\newcommand\be{\beta}

\newcommand\de{\delta}

\newcommand\si{\sigma}

\newcommand\la{\langle}
\newcommand\ra{\rangle}

\newcommand\pa{\partial}
\newcommand\ua{\uparrow}
\newcommand\da{\downarrow}

\begin{document}

\title{Confining electrons on a topological insulator surface using 
potentials and a magnetic field}
% Effects of a barrier on the surface of a topological insulator} 

\author{Ranjani Seshadri and Diptiman Sen}
\affiliation{\small{Centre for High Energy Physics, Indian Institute of 
Science, Bangalore 560 012, India}} 

\date{\today}

\begin{abstract}
We study the effects of extended and localized potentials and a magnetic 
field on the Dirac electrons residing at the surface of a three-dimensional 
topological insulator. We use a lattice model to numerically study the various
states; we show how the potentials can be chosen in a way which effectively 
avoids the problem of fermion doubling on a lattice. We show that extended 
potentials of different shapes can give rise to states which propagate freely 
along the potential but decay exponentially away from it. For an infinitely
long potential barrier, the dispersion and spin structure of these states 
are unusual and these can be varied continuously by changing the barrier 
strength. In the presence of a magnetic field applied perpendicular to the 
surface, these states become separated from the gapless surface states by a 
gap, thereby giving rise to a quasi-one-dimensional system. Similarly, a 
magnetic field along with a localized potential can give rise to exponentially
localized states which are separated from the surface states by a gap and 
thereby form a zero-dimensional system. Finally, we show that a long barrier 
and an impurity potential can produce bound states which are localized at the 
impurity, and an ``L"-shaped potential can have both bound states at the 
corner of the ``L" and extended states which travel along the arms of the 
potential.
\end{abstract}

\maketitle

\section{Introduction}
Topological insulators (TIs) are materials which have gapped states in the 
bulk and gapless states on the boundaries which are protected by time-reversal
symmetry~\cite{tirev1,tirev2}. These materials have been studied both 
theoretically~\cite{zhang1,kane1,kane2,teo,qi} and 
experimentally~\cite{hasan1,exp2,exp1,hasan2} for a number of years.
Three-dimensional TIs such as $\rm Bi_2 Se_3$, $\rm Bi_2 Te_3$ and 
$\rm Sb_2 Te_3$ have 
surfaces on which there is a single species of gapless states which is 
governed by the Dirac equation~\cite{hasan1,exp2,hasan2}. A number of 
interesting properties of the surface states of a TI have been 
studied~\cite{kane4,tirev1,tirev2,fu,been1,tanaka1,tanaka2,mondal1,hasan2,
burkov,garate,zyuzin,das1}. Junctions of different surfaces of TIs, 
sometimes separated by a geometrical step or a magnetic domain 
wall~\cite{taka,deb1,wickles,biswas,alos,sitte,zhang3,apalkov,habe}, junctions 
of surfaces of a TI with normal metals or magnetic materials~\cite{modak} or
superconductors~\cite{soori}, and polyhedral surfaces~\cite{ruegg} have been
investigated. The effects of finite sizes~\cite{shen,linder,egger,shenoy,
pertsova,neupane} and different 
orientations~\cite{silvestrov,zhang3,barreto,rao,deb2}, and transport around 
different surfaces of a TI in the presence of a magnetic field~\cite{peng} 
have been studied. The effect of a periodically varying one-dimensional
potential and a magnetic field on the spectrum of electrons on the surface of
$\rm Bi_2 Te_3$ has been studied in Ref.~\onlinecite{okada}.

It is known analytically that an infinitely long $\de$-function potential 
barrier running along the $x$ axis applied to the $x-y$ surface of a TI
gives rise to states which propagate as plane waves along the barrier and 
decay exponentially away from it~\cite{deb1}. However, there is no energy 
gap between the states produced by the potential barrier and the gapless
surface states which exist far from the potential. As a result, the former
states are not robust against disorder; even a weak disorder can produce a 
transition between these states and the gapless surface states. On the other 
hand, a potential localized in some region does not produce any localized 
states. If we now apply a Zeeman field perpendicular to the surface, the 
surface states get gapped out. It is then possible that localized potentials 
will also produce localized states, and that the states produced by various
potentials (either extended or localized) will lie in the gap of the surface
states and will therefore be stable against weak disorder. Hence, 
we can produce systems confined to one or zero dimensions (resembling quantum 
wires and dots) which may be useful for various practical applications.

In this paper, we use a lattice model to study the states produced by
a combination of non-magnetic potentials and a perpendicular magnetic field 
which is Zeeman coupled to the spin of the electrons. A lattice model allows 
us to numerically study the effects of potentials of any magnitude and shape. 
The plan of our paper is as follows. In Sec. II, we review the 
Dirac Hamiltonian in the continuum and its symmetries in the presence of a 
potential and a magnetic field. In Sec. III we discretize the Hamiltonian
using a square lattice and we discuss the fermion doubling problem that arises
for Dirac electrons. In Sec. IV, we numerically study the spectrum of 
electrons in the presence of an infinitely long potential barrier which is 
taken to have a Gaussian profile in the transverse direction. Since this 
system has translation invariance along one direction, the momentum along that 
direction is a good quantum number and can be used to reduce it to a 
one-dimensional problem. In the absence of a magnetic field, the dispersion 
of the states propagating along the barrier (henceforth called edge states) is
qualitatively found to be of the form $E = v|k_x|$, where $k_x$ is the momentum
along the barrier, if $|k_x|$ is much smaller than the inverse lattice spacing.
The expectation value of the spin, $\la {\vec \si} \ra$,
of the edge states lies in the $y$ direction. The velocity $v$ of the 
the edge states is smaller than that of the surface states and it 
can be varied by changing the strength of the potential barrier. 
The velocity $v$ is found to be very small for a particular 
value of the potential barrier, giving rise to an almost flat band near 
$E=0$. The wave function of the edge states decays exponentially
away from the potential barrier; the decay length is found to be inversely
proportional to $|k_x|$. Hence the edge states will cease to exist when 
the decay length becomes comparable to the size of the system.
When a Zeeman field is applied in the $z$ direction, the surface 
states become gapped but the edge states do not. Further, the edge state
now exists even for $k_x = 0$, and their dispersion can be controlled by
the potential barrier. The edge states then define a tunable one-dimensional 
system which is separated from the surface states by a gap. In Sec. V, we 
study the effects of a variety of potentials with two-dimensional profiles. 
We first consider a potential localized in some region. In the absence of a 
Zeeman field there are no localized states, but when a Zeeman field is turned 
on, we find that there exponentially localized states can appear if the 
potential is strong enough. Next, we study
a combination of a long potential barrier, a localized potential and a 
magnetic field; we find that states can appear which are bound to the 
localized potential. Finally, we study what happens if there is an ``L"-shaped
potential consisting of two infinitely long arms meeting at a corner and
a magnetic field. We find that there can be both states bound to the corner
of the ``L" as well as scattering states which propagate along the arms.
In Sec. VI, we summarize our main results and 
describe some ways of experimentally testing these results.

\section{Surface Hamiltonian}
The surface states of a three-dimensional TI are governed by the massless 
Dirac equation. The form of the Dirac equation depends on the orientation
of the surface~\cite{silvestrov,zhang3,barreto,rao,deb2}; 
the simplest form appears when the surface is given by the $x-y$ plane. 
We will also be interested in the effects of a scalar potential 
$V(x,y)$ and a uniform magnetic field $\vec B$ which only has a Zeeman 
coupling to the electrons. Including these terms, the two-component wave 
function $\psi (x,y) e^{-iEt}$ of an energy eigenstate satisfies the equation 
\beq [-i v_F (\si^x \pa_y - \si^y \pa_x) + V - \frac{g\mu}{2} {\vec \si} \cdot
{\vec B}] \psi = E \psi, \label{ham1} \eeq
where $v_F$, $\mu$ and $g$ denote the Fermi velocity, the Bohr magneton, and 
the gyromagnetic ratio respectively. (We will set $\hbar = 1$ in this paper).

{\bf Spin-momentum locking:}
If both $V$ and $\vec B$ are absent, the solutions of Eq.~\eqref{ham1} have 
momenta ${\vec k} = (k_x,k_y)$ and energies $E_\pm = \pm v_F \sqrt{k_x^2 +
k_y^2}$. The wave functions are given by $\psi e^{i(k_x x + k_y y -Et)}$, 
where \bea \psi_+ &=& \frac{1}{\sqrt 2} ~\left( \begin{array}{c}
1 \\
\frac{k_y - i k_x}{E} \end{array} \right) ~~{\rm for}~~ E_+, \non \\
\psi_- &=& \frac{1}{\sqrt 2} ~\left( \begin{array}{c}
1 \\
- ~\frac{k_y - i k_x}{E} \end{array} \right) ~~{\rm for}~~ E_-. \label{soln}
\eea
Upon calculating the expectation values $\la \si^x \ra$, $\la \si^y \ra$
and $\la \si^z \ra$, we find that the direction of spin is perpendicular to 
both $\hat z$ and ${\hat k} = {\vec k}/|{\vec k}|$, namely, $\la {\vec \si} 
\ra = {\hat k} \times {\hat z}$ and $- {\hat k} \times {\hat z}$ for $E_+$ 
and $E_-$ respectively. This property of the surface states is called
spin-momentum locking.

If we now turn on a magnetic field perpendicular to the surface, ${\vec B} 
= B_z {\hat z}$, the states with momentum $(k_x,k_y)$ will have energies 
$E_\pm = \pm \sqrt{v_F^2 (k_x^2 + k_y^2) + (g\mu B_z/2)^2}$; hence there will 
be a gap of $|g\mu B_z|$ at ${\vec k} = 0$. Further, these states have a 
non-zero value of $\la \si^z \ra$.

{\bf Effect of a potential:}
Let us now turn on a potential $V (x,y)$ but no magnetic field ${\vec B}$.
Then Eq.~\eqref{ham1} takes the form
\beq [v_F (-i\si^x \pa_y + i \si^y \pa_x) + V(x,y)] \psi = E \psi.
\label{ham2} \eeq
Eq.~\eqref{ham2} has the following symmetries.

\noi (i) Time-reversal symmetry $\cal T$: Eq.~\eqref{ham2} remains invariant
if we complex conjugate all numbers, and transform and $\psi (x,y) \to 
\si^y \psi^* (x,y)$. Since $\si^{y*} = - \si^y$, we have ${\cal T}^2 = -I$;
this implies that every energy level must have a two-fold degeneracy.

\noi (ii) Parity symmetry $\cal P$: If the potential is invariant under 
reflection in $y$, i.e., $V(x,-y) = V(x,y)$, we have a symmetry $\cal P$
under which $\psi (x,y) \to \psi^* (x,-y)$.

\noi (iii) $\pi$-rotation symmetry $\cal R$: If the potential is invariant 
under a $\pi$ rotation about the $\hat z$ axis, i.e., $V(-x,-y)= V(x,y)$, we 
have a symmetry $\cal R$ under which $\psi (x,y) \to \si^z \psi (-x,-y)$. 

If a magnetic field is applied in the $z$ direction, time-reversal symmetry
is broken but $\cal P$ and $\cal R$ hold if $V$ has both parity and 
rotational symmetries.

%\noi (ii) Parity symmetries: Assuming that $V(-y) = V(y)$, there are two 
%parity symmetries, namely, ${\cal P}_x$ under which we transform $\psi(k_x,y) 
%\to \si^x \psi(-k_x,y)$, and ${\cal P}_y$ under which $\psi(k_x,y) \to \si^y 
%\psi(k_x,-y)$. These two transformations anticommute: ${\cal P}_x 
%{\cal P}_y = - {\cal P}_y {\cal P}_x$. The existence of two anticommuting
%symmetries implies that for each momentum $k_x$, the energy $E$ has a
%two-fold degeneracy. This can be proved by contradiction: if $\psi$ is a
%non-degenerate eigenstate, we must have ${\cal P}_x \psi = \lam_x \psi$ and
%${\cal P}_y \psi = \lam_y \psi$, where $\lam_x, \lam_y$ are some numbers (equal
%to $\pm 1$ since ${\cal P}_x^2 = {\cal P}_y^2 = I$). This will contradict
%${\cal P}_x {\cal P}_y \psi = - {\cal P}_y {\cal P}_x \psi$. Finally, we 
%observe that $i{\cal P}_x {\cal P}_y$ is also a symmetry of the Hamiltonian. 
%This transforms $\psi(k_x,y) \to -\si^z \psi(-k_x,-y)$ which corresponds to 
%a $\pi$ rotation in the $x-y$ plane. (Within the two-dimensional subspace of 
%degenerate eigenstates, the three operators ${\cal P}_x$, ${\cal P}_y$ and 
%$i{\cal P}_x {\cal P}_y$ can be mapped to three Pauli matrices).

%\noi (iii) Complex conjugation symmetry $\cal C$: Eq.~\eqref{ham2} remains
%invariant if we complex conjugate all numbers and transform $\psi (k_x,y) \to 
%\si^z \psi^* (k_x,y)$.

In Ref.~\onlinecite{deb1}, the effect of a $\de$-function potential barrier,
$V(y) = V_0 \de (y)$, was studied analytically. It was shown that this can
give rise to states which propagate as plane waves in the $x$ direction
and decay exponentially as one moves away from $y=0$. In the next section,
we will consider the effect of more complicated potentials as well as a 
magnetic field applied in the $z$ direction.

\section{Lattice Model and Fermion Doubling}

For a general form of the potential $V(x,y)$, for example in the 
presence of impurities, it is not possible to find the energy spectrum
and wave functions analytically and one has to resort to a numerical 
solution. For this purpose, we assume the $x-y$ plane to be a lattice of 
discrete points $\{ n_x, n_y \}$ and Eq.~\eqref{ham1} becomes,
\bea && - ~\frac{1}{2} ( \be_{n_x+1,n_y}- \be_{n_x-1,n_y}) ~-~ 
\frac{i}{2} ( \be_{n_x,n_y+1}-\be_{n_x,n_y-1} ) \non \\
&& +~ (V_{n_x,n_y}+B) \al_{n_x,n_y} ~=~ E \al_{n_x,n_y}, \label{eq:lat1} \\
&& \non \\
&& \frac{1}{2} ( \al_{n_x+1,n_y}- \al_{n_x-1,n_y} ) ~-~ \frac{i}{2} ( 
\al_{n_x,n_y+1}-\al_{n_x,n_y-1} ) \non \\
&& + (V_{n_x,n_y}-B)\be_{n_x,n_y} ~=~ E \be_{n_x,n_y}, \label{eq:lat2} \eea
where $\al_{n_x,n_y}$ and $\be_{n_x,n_y}$ respectively denote the wave 
functions of spin-$\ua$ and spin-$\da$ electrons at the site $\{n_x,n_y\}$, 
we have assumed a magnetic field $\vec B =B_z \hat z$ with $B= -g \mu B_z / 2$.
In our numerical calculations, we will work in units in which the velocity 
$v_F$ and lattice spacing $a$ are both equal to unity; at the end of Sec. V
we will restore all the physical units for a topological insulator like
$\rm Bi_2 Se_3$.

\begin{figure}[htb]
\begin{center}
\includegraphics[width=6cm]{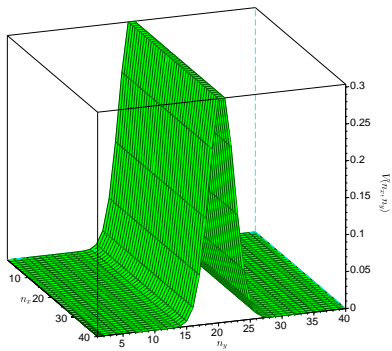}
\end{center}
\caption{Surface plot of a potential barrier which is a Gaussian in one
direction with width 2 and $V_b = \pi/2$, as described in 
Eq.~\eqref{eq:barr}.} \label{fig:V} \end{figure}

{\bf Fermion doubling:}
Eqs.~(\ref{eq:lat1}-\ref{eq:lat2}) suffer from the problem of ``fermion 
doubling". To see this in the simplest possible way, consider the case when 
both the potential and the magnetic field are absent, i.e., $V=0$ and $B=0$. 
Translational symmetry along both $x$ and $y$ directions allows the solution
\bea \begin{pmatrix} \al_{n_x,n_y} \\ 
\be_{n_x,n_y} \end{pmatrix} &=& \begin{pmatrix} \al \\ 
\be \end{pmatrix} e^{i\left( k_x n_x + k_y n_y\right)}. \eea
This gives
\bea [ \sin (k_y) - i \sin (k_x) ] ~\be &=& E \al, \non \\
{} [ \sin (k_y) + i \sin (k_x) ] ~\al &=& E \be, \eea
which leads to the dispersion relation
\beq E = \pm \sqrt{\sin^2 (k_x)+ \sin^2(k_y)}. \label{eq:lin_disp} \eeq
Clearly this vanishes at four points in the Brillouin zone lying at $(0,0)$,
$(0,\pi)$, $(\pi,0)$ and $(\pi,\pi)$, giving rise to four Dirac cones, in 
contrast to the continuum theory which only has one Dirac cone at $(0,0)$. 

One way to avoid this problem is to add the Wilson term; this is proportional 
to $\si^z$ and it adds $w \big[ 2 -\cos(k_x)-\cos(k_y)) \big] \al_{n_x,n_y}$ 
in Eq.~\eqref{eq:lat1} and $- w \big[ 2 -\cos(k_x)-\cos(k_y)) \big] 
\be_{n_x,n_y}$ in Eq.~\eqref{eq:lat2} on the left hand sides. The dispersion 
relation now becomes,
\bea E &=& \pm ~[~\sin^2(k_x) ~+~ \sin^2(k_y) \non \\
&& ~~~~+ w^2 (2-\cos(k_x) - \cos(k_y))^2 ~]^{1/2}. \eea
This reduces to Eq.~\eqref{eq:lin_disp} in the low momentum limit, but it 
does not vanish near the boundaries of the Brillouin zone where $k_x$ or 
$k_y$ approaches $\pm \pi$. We thus recover a system with only one Dirac 
cone lying at $(0,0)$.

Since the Wilson term is proportional to $\si^z$, it looks like a magnetic 
field in the $z$ direction; hence it breaks some symmetries such as 
time-reversal symmetry and gives rise to various spurious effects. We can 
avoid working with a Wilson term if we can 
ensure that the wave functions that we are studying only have momentum
components lying close to $(k_x,k_y) = (0,0)$. This will be true if
our potentials are sufficiently smooth so that their Fourier 
components rapidly approach zero as we move away from $(0,0)$, and
if our system sizes are large (since the smallest possible momentum is
inversely proportional to the system size). Our numerical results presented 
below will show that choosing smooth potentials enables us to effectively 
avoid the fermion doubling problem even without adding a Wilson term.

{\bf Bound states and inverse participation ratio:} 
In our numerical studies, we will be specially interested in states 
which are localized in certain regions of space. We will refer to all
such states as bound states for simplicity. Bound states can be identified
most easily by inverse participation ratios (IPR) of all the energy 
eigenstates. Let $\psi_{i;n_x,n_y}$ be the value of the wave 
function at the lattice site $(n_x,n_y)$ with the $i^{th}$ energy eigenvalue 
$E_i$. The normalization condition implies that 
\beq \sum_{n_x,n_y} ~\Big|\psi_{i;n_x,n_y}\Big|^2 ~=~ 1. \eeq
We now define
\beq \Big(IPR\Big)_i = \sum_{n_x,n_y} ~\Big|\psi_{i;n_x,n_y}\Big|^4. \eeq
The more localized the wave function of a particular state is, the higher will
its IPR be. This can be understood from following example. If a normalized 
wave function has the form
\beq \psi_{n_x,n_y} ~\sim~ \frac{e^{-(n_x^2 + n_y)^2/\xi^2}}{\xi}, \eeq
its IPR will be proportional to $1/\xi^2$. Hence the state with the 
smallest width $\xi$ will have the largest IPR.

\section{Numerical results in one dimension}

We first study the energy spectrum in the presence of a potential $V$ which
is only a function of $n_y$ on a lattice. The spectrum can be found assuming 
the wave function to have a momentum $k_x$ along the $x$ direction; this 
reduces it to a one-dimensional problem involving $\psi = (\al_{n_y}, ~
\be_{n_y})^T e^{i(k_x x - Et)}$. The eigenvalue problem is given by
\bea && -~ i \sin(k_x) ~\be_{n_y} ~-~ \frac{i}{2} ( \be_{n_y+1} -
\be_{n_y-1} ) ~+~ V_{n_y} \al_{n_y} \non \\
&& =~ E \al_{n_y}, \non \\
&& i \sin(k_x) ~\al_{n_y} ~-~ \frac{i}{2} ( \al_{n_y+1}-\al_{n_y-1} ) ~
+~ V_{n_y} \be_{n_y} \non \\
&& =~ E \be_{n_y}. \label{eq:1D} \eea

We take the potential to be a barrier which is a Gaussian with an integrated 
weight of 
\beq V_{n_y} ~=~ \frac{V_b}{\si \sqrt{2\pi}} ~e^{-(n_y - n_0)^2/(2\si^2)}, 
\label{eq:barr} \eeq
with maximum value $V_b/(\si \sqrt{2\pi})$ and width $\si$. In our 
calculations, we will set the width $\si =2$ and vary $V_b$. We will take the 
Gaussian to be centered at $n_y = n_0 = 151$ for a system with 301 sites in 
the $y$ direction. For $\si=2$, the Fourier transform ${\tilde V}_{k_y} = 
\sum_{n_y} e^{-ikn_y} V_{n_y}$ decreases rapidly as we go away from $k_y =0$ 
and is very small at $k_y =\pi$. Using the Poisson resummation formula, we 
obtain
\beq \Big| \frac{\tilde V_{k_y}}{\tilde V_{k_y=0}} \Big| ~=~ \frac{
\sum_{n=-\infty}^\infty ~e^{-2 \pi^2 \si^2 [n - k_y /(2\pi)]^2}}{
\sum_{n=-\infty}^\infty ~e^{-2 \pi^2 \si^2 n^2}}, \label{fourier} \eeq
For $\si =2$, we find that Eq.~\eqref{fourier} is very well approximated by 
the Gaussian $e^{-2k_y^2}$ for $k_y$ lying in the range $[-\pi, \pi]$. At
$k_y = \pi$, $|{\tilde V}_\pi /{\tilde V}_0| \simeq 3 \times 10^{-9}$ which 
is extremely small. Hence a Gaussian potential with width 2 is sufficiently 
smooth so that states near $k_y=\pi$ make very little contribution to the 
bound states. Indeed, as mentioned below, we find numerically that the wave 
functions of the energy eigenstates are quite smooth, with period 2 
oscillations (corresponding to components of $k_y$ close to $\pi$) being 
rather small.

\begin{figure}[h!]
\begin{center}
%\subfigure[]{\includegraphics[width=4.2cm]{EvsK_B_0_Vo_piby4.eps}}
\subfigure[]{\includegraphics[width=4.2cm]{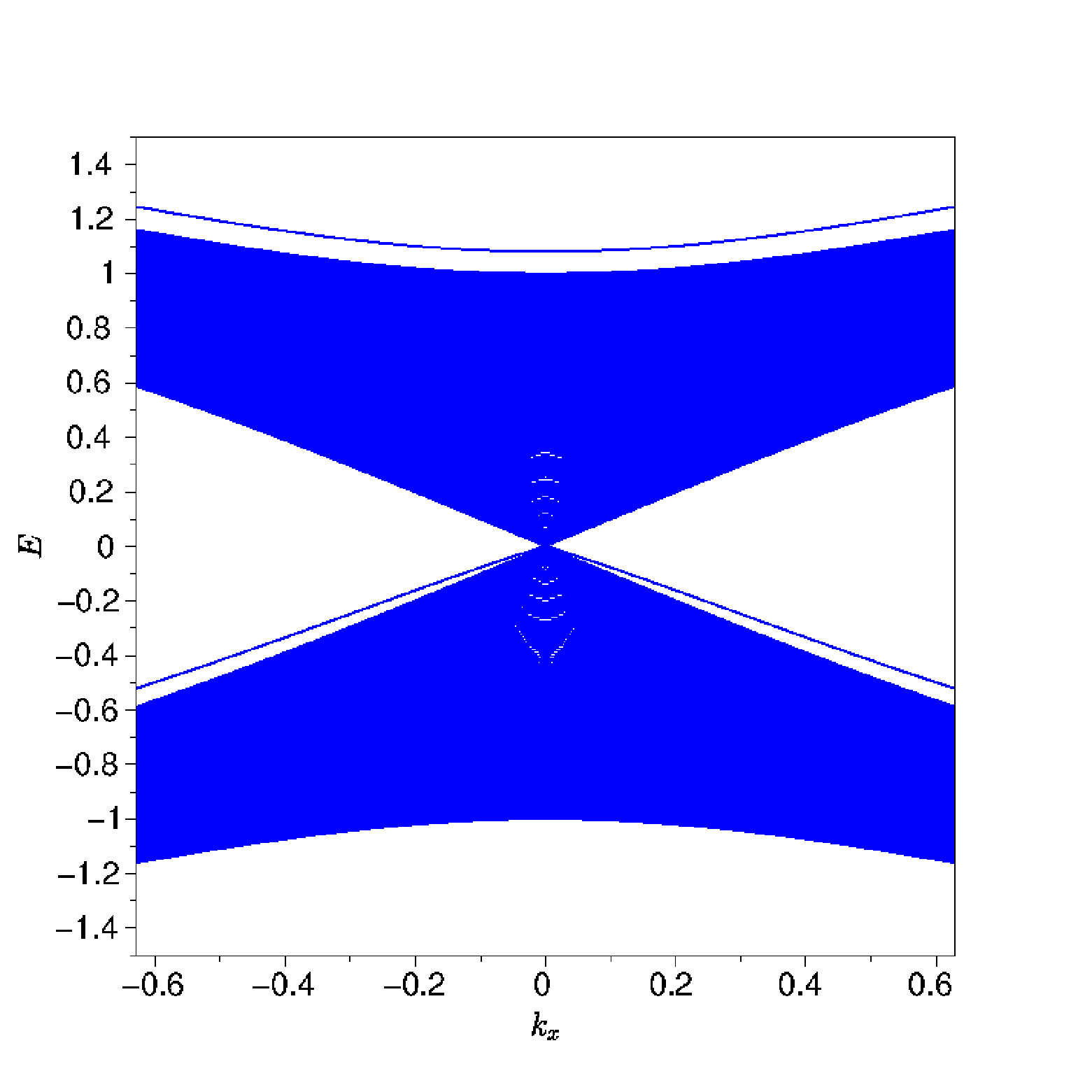}}
%\subfigure[]{\includegraphics[width=4.2cm]{fig02.png}}
%\subfigure[]{\includegraphics[width=4.2cm]{EvsK_B_0_Vo_piby2.eps}}
\subfigure[]{\includegraphics[width=4.2cm]{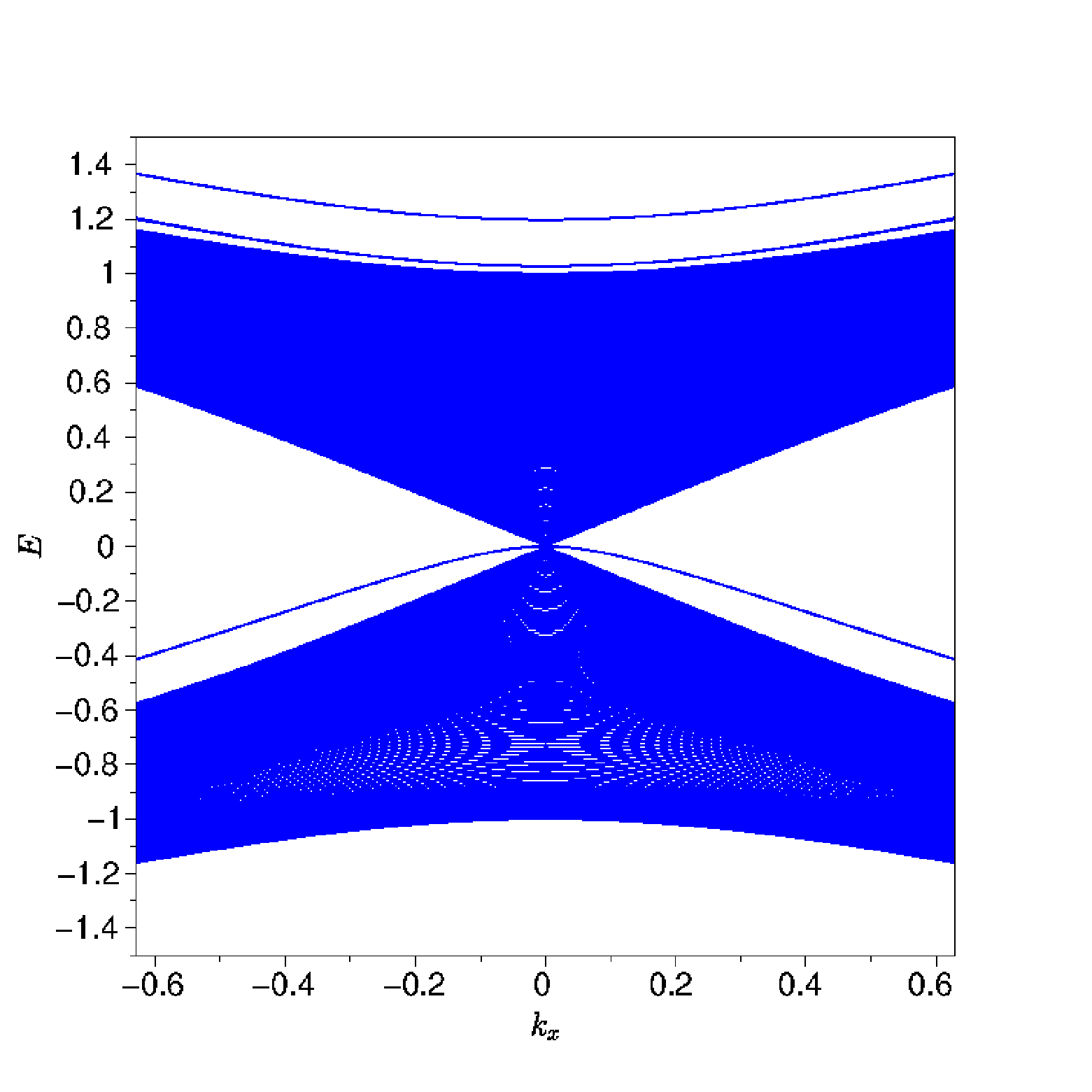}}
%\subfigure[]{\includegraphics[width=4.2cm]{fig03.png}}
\\
%\subfigure[]{\includegraphics[width=4.2cm]{EvsK_B_piby20_Vo_piby4.eps}}
\subfigure[]{\includegraphics[width=4.2cm]{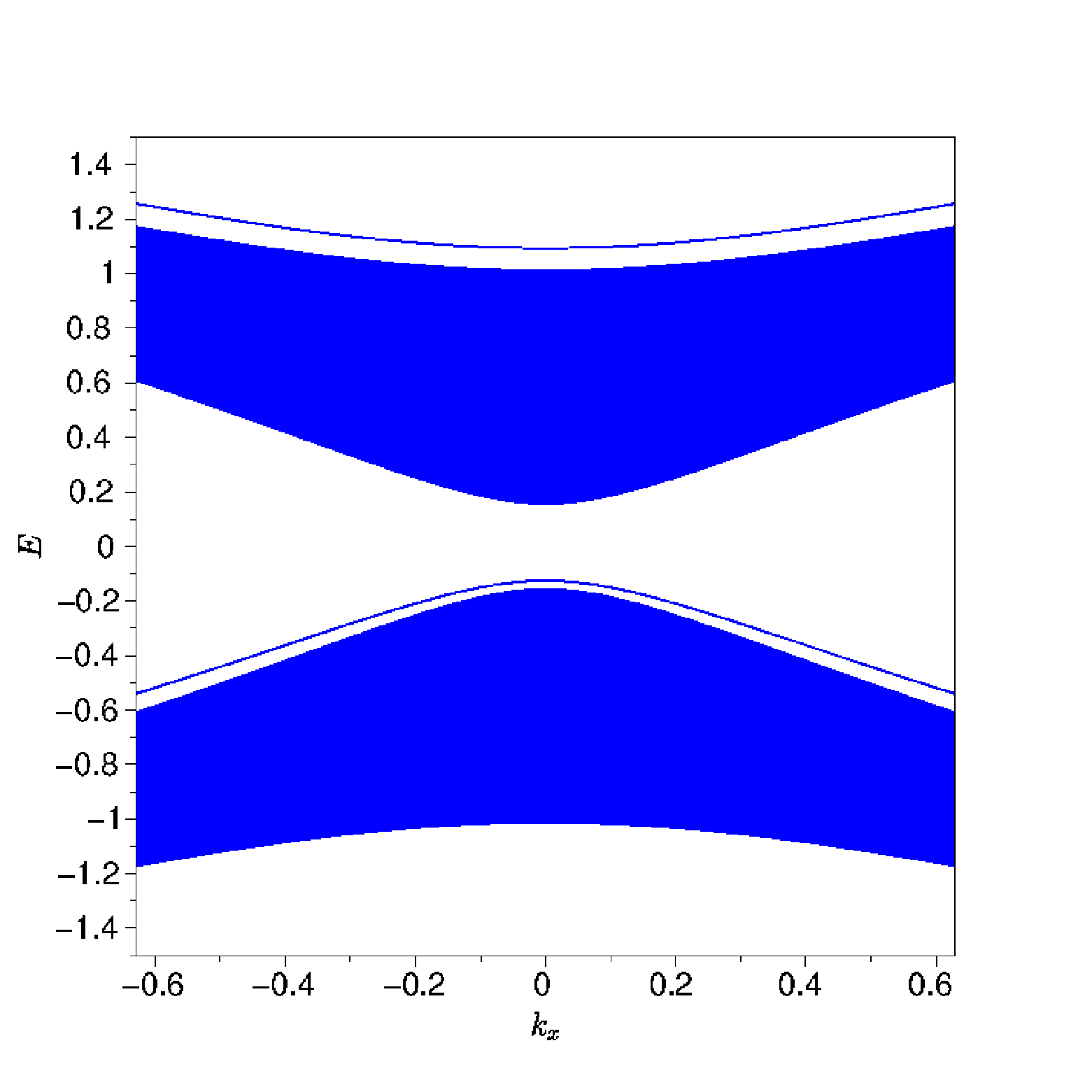}}
%\subfigure[]{\includegraphics[width=4.2cm]{fig04.png}}
%\subfigure[]{\includegraphics[width=4.2cm]{EvsK_B_piby20_Vo_piby2.eps}}
\subfigure[]{\includegraphics[width=4.2cm]{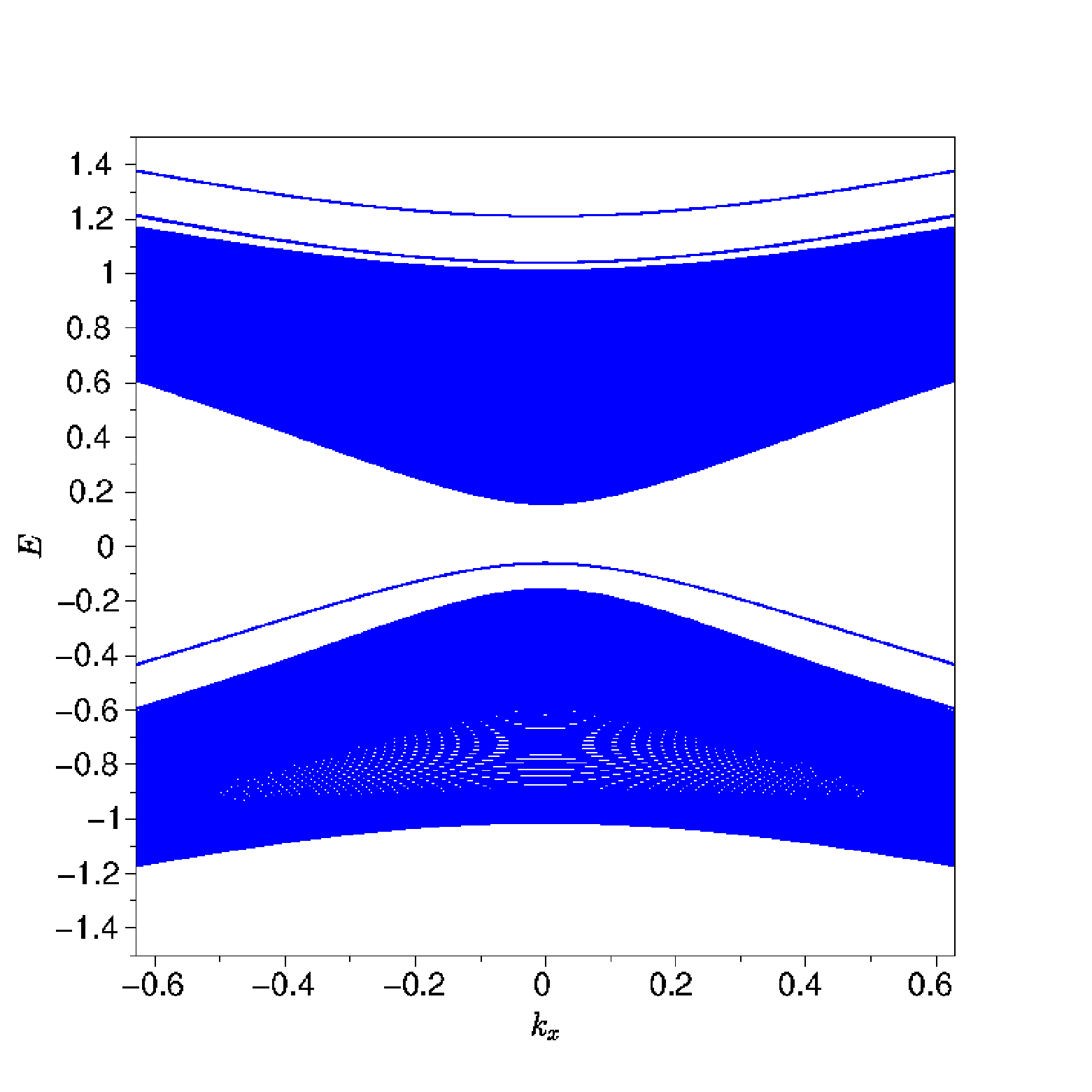}}
%\subfigure[]{\includegraphics[width=4.2cm]{fig05.png}}
\end{center}
\caption{Energy dispersion of barrier states for (a) $V_b=\pi/4$ and $B=0$,
(b) $V_b=\pi/2$ and $B=0$, (c) $V_b=\pi/4$ and $B=\pi/20$, and (d) $V_b=
\pi/2$ and $B=\pi/20$, for $k_x$ lying in the range $[-\pi/5,\pi/5]$.} 
\label{fig:disp} \end{figure}

{\bf Bound states:} We first consider the case when no magnetic field is 
applied. As $V_b$ is increased from zero, we find that a set of bound states 
appears which are separated from the plane wave surface states which have the 
gapless spectrum $E = \pm \sqrt{k_x^2 + k_y^2}$. The new states are plane 
waves along the $x$ direction and decay exponentially as one moves away from 
the centre of the barrier. The energy $E$ of these states is a function
of $|k_x|$; this is a consequence of both the symmetries $\cal P$ and $\cal R$
mentioned earlier. The ratio $dE/d|k_x|$ close to $k_x = 0$ varies with the 
potential strength $V_b$; it has a value of $-v_F = -1$ close to $V_b=0$
and increases as $V_b$ is increased, becoming almost zero around $V_b=\pi/2$. 
This is illustrated in the top panels of Fig.~\ref{fig:disp}, with $V_b=\pi/4$ 
and $\pi/2$ in Figs.~\ref{fig:disp} (a) and (b) respectively. ($E/|k_x|$ 
becomes positive when $V_b$ is increased beyond $\pi/2$). 

We thus see that an almost flat band can be produced by tuning the barrier 
strength $V_b$. For such a band, states with different momenta can be 
superposed suitably to give any wave function that one chooses, and all such 
states will have almost the same energy. Further, such states will move
only slowly in time since the group velocity $dE/dk_x$ is close to zero.
\vskip .2cm

The states bound to the potential barrier are degenerate with the surface
states. In the presence of scattering (induced by, say, impurities which
may be present close to the barrier), an electron occupying a bound states 
can scatter into a surface state. The bound states can be made immune to such
scattering by introducing a magnetic field. 
Figs.~\ref{fig:disp} (c) and (d) show the bound states when a magnetic
field given by $B=\pi/20$ is introduced. This opens a gap of $2B$ in the
spectrum of the surface states, and the bound states which lie within this
gap can be expected to be robust against scattering from impurities.

We note that Figs.~\ref{fig:disp} show some additional sets of states near
above the top of the band of surface states. The wave functions of these 
states oscillate rapidly on the scale of a lattice spacing (they have momentum
components close to $\pi$); hence, they are lattice artifacts and have no 
counterparts in the continuum limit of the model.

\begin{figure}[htb]
\begin{center}
%\subfigure[]{\includegraphics[width=4.2cm]{Prob_B_0_Vo_piby2_kx_mpiby10.eps}}
\subfigure[]{\includegraphics[width=4.2cm]{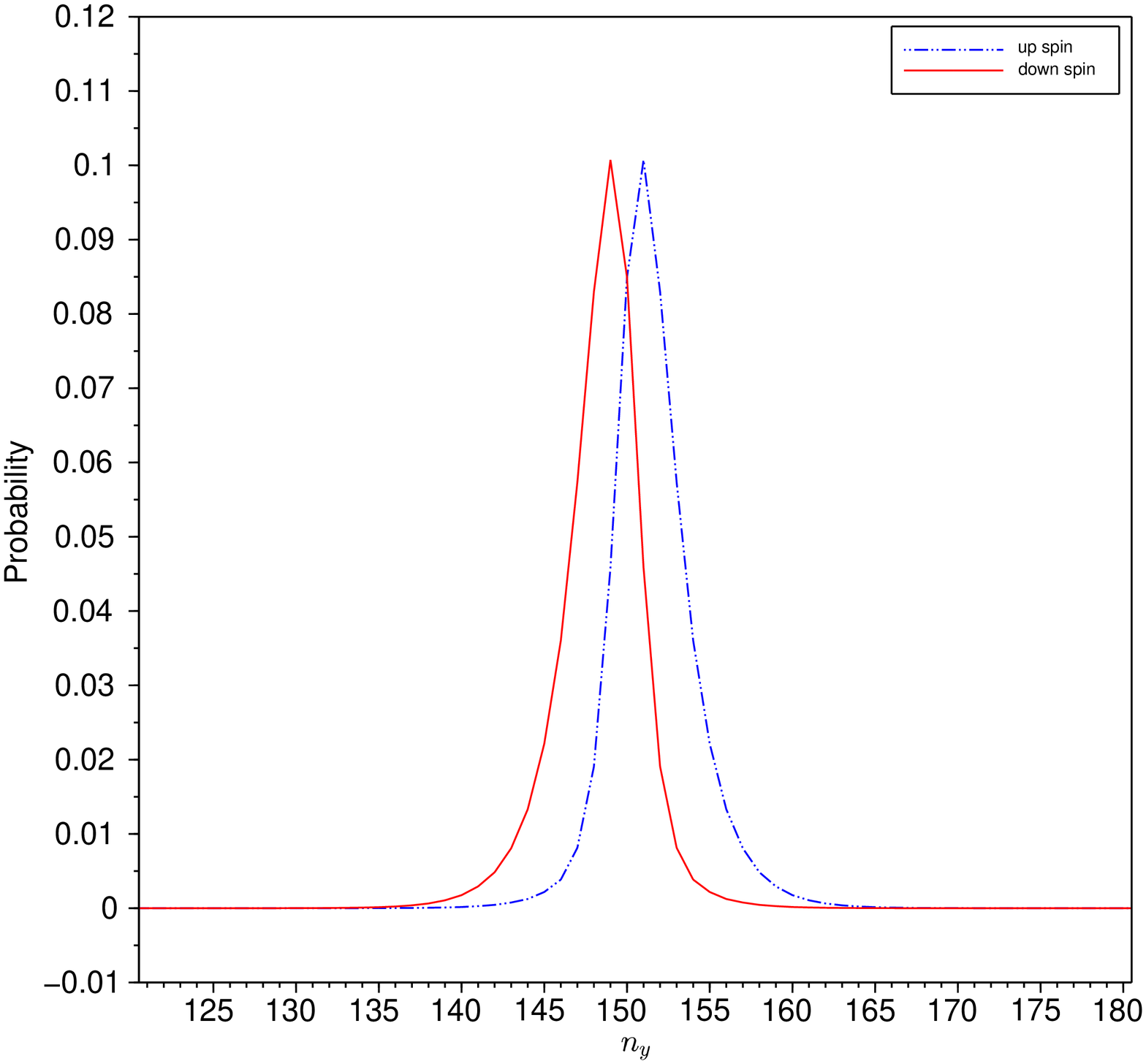}}
%\subfigure[]{\includegraphics[width=4.2cm]{Prob_B_piby20_Vo_piby2_kx_mpiby10.eps}}
\subfigure[]{\includegraphics[width=4.2cm]{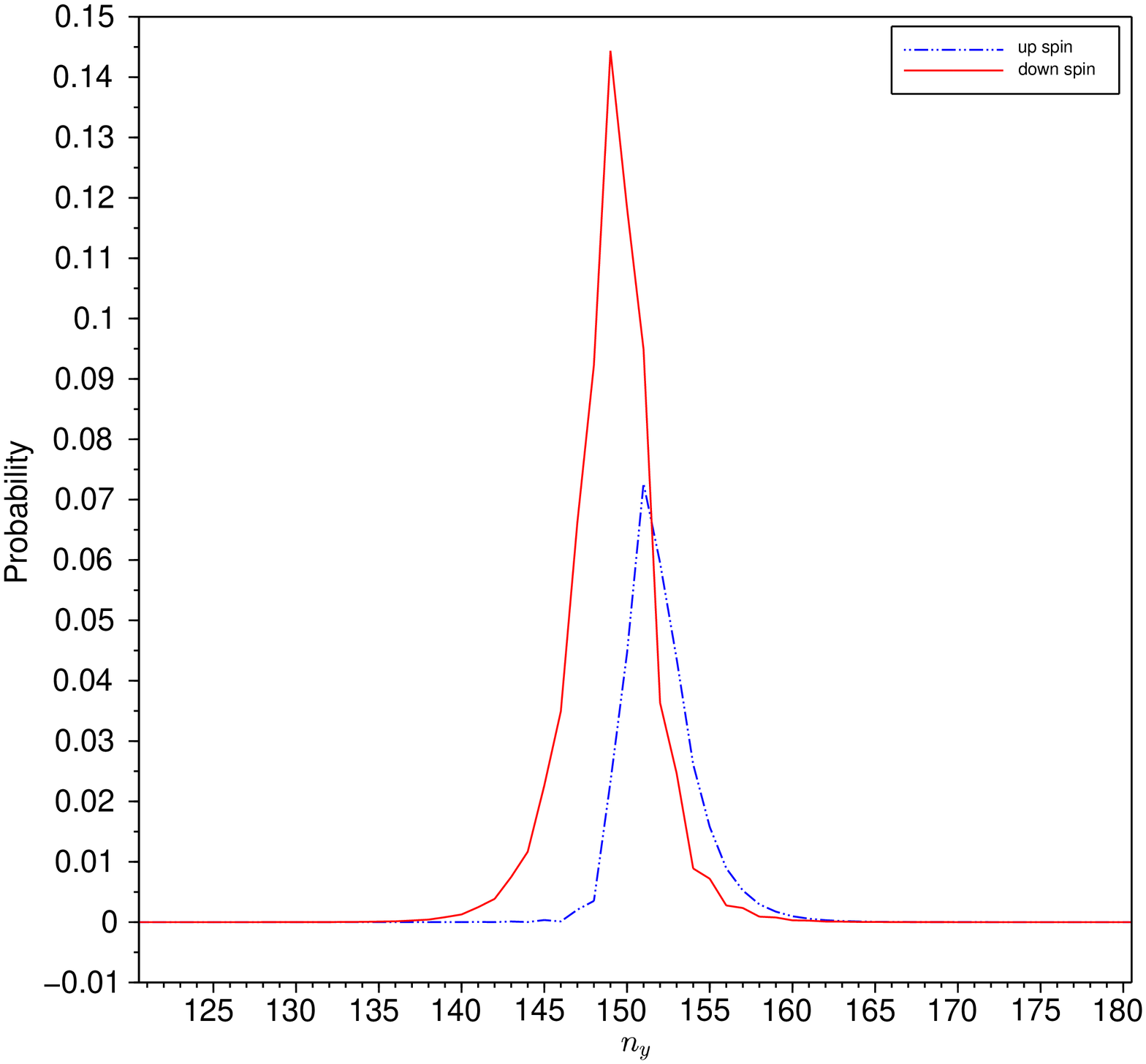}}
\\
%\subfigure[]{\includegraphics[width=4.2cm]{Prob_B_0_Vo_piby2_kx_0.eps}}
\subfigure[]{\includegraphics[width=4.2cm]{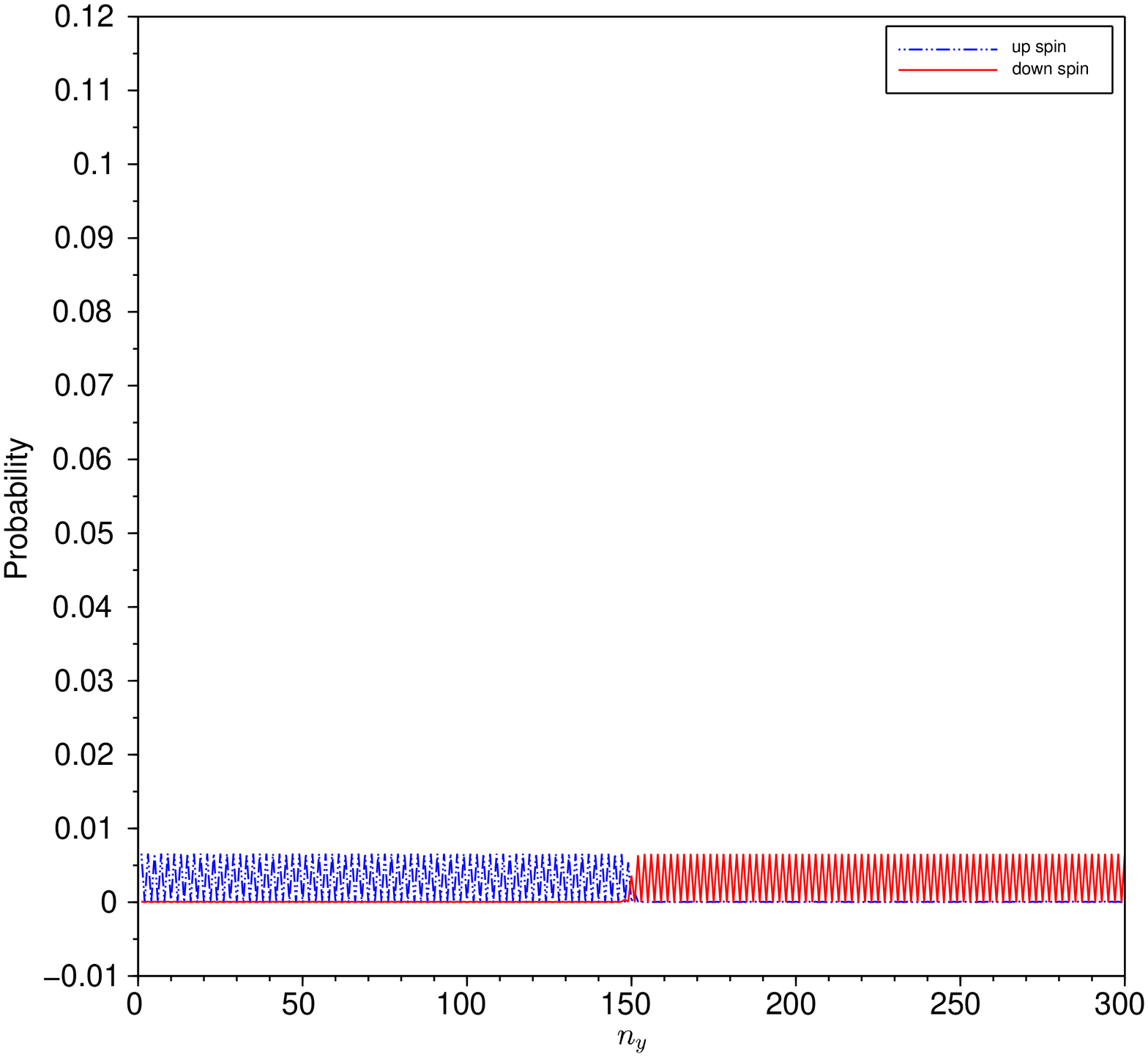}}
%\subfigure[]{\includegraphics[width=4.2cm]{Prob_B_piby20_Vo_piby2_kx_0.eps}}
\subfigure[]{\includegraphics[width=4.2cm]{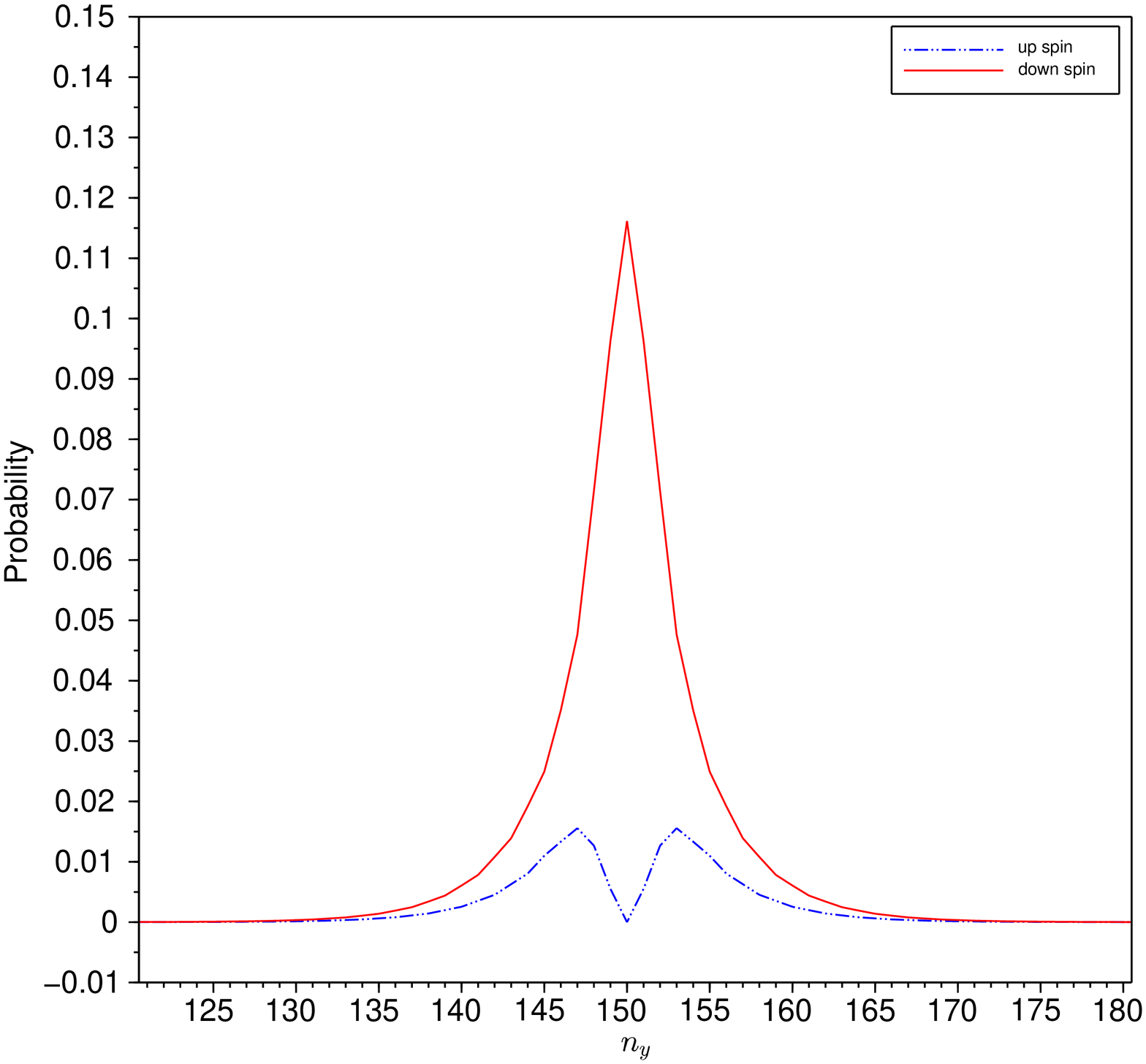}}
\\
%\subfigure[]{\includegraphics[width=4.2cm]{Prob_B_0_Vo_piby2_kx_piby10.eps}}
\subfigure[]{\includegraphics[width=4.2cm]{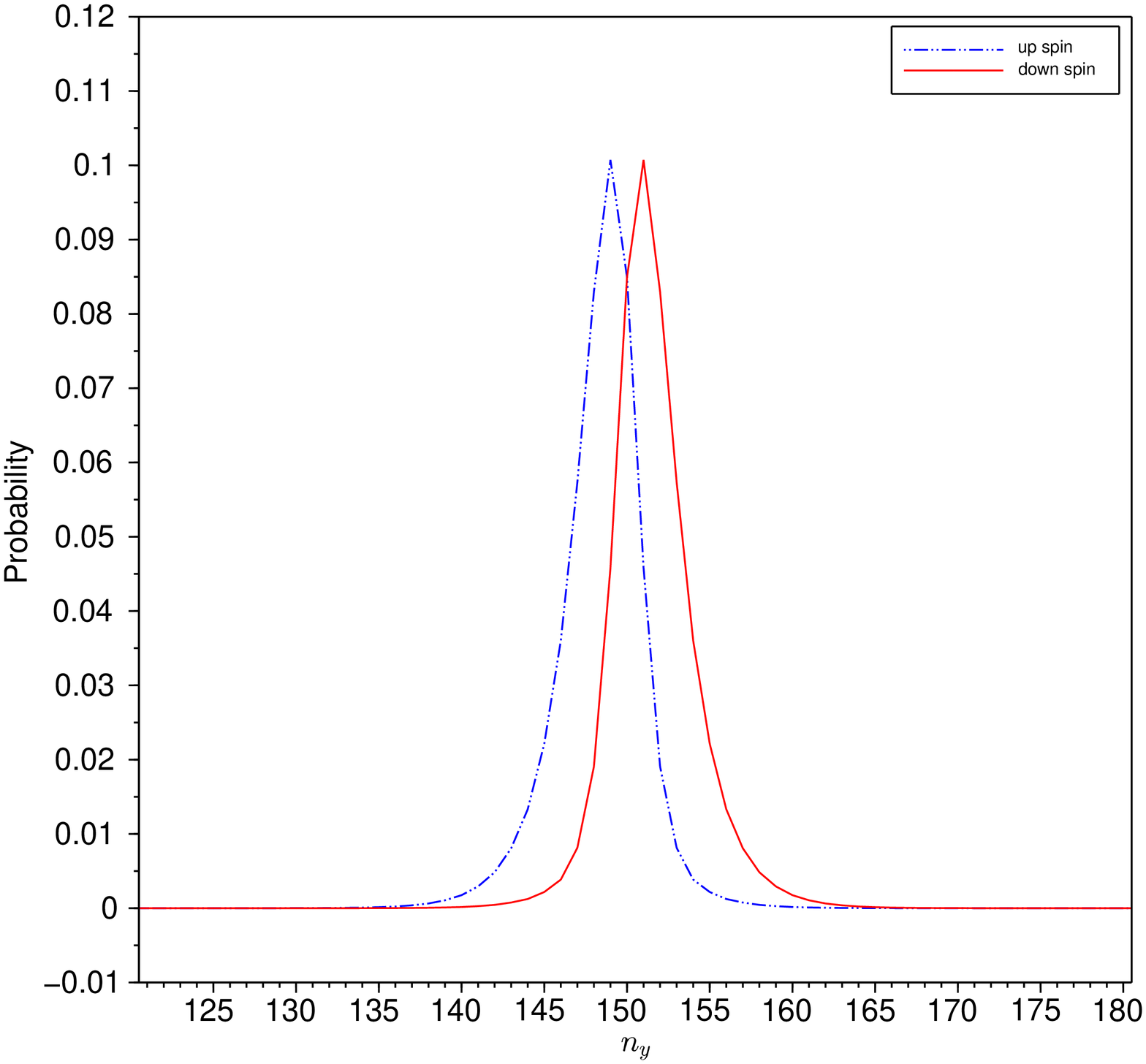}}
%\subfigure[]{\includegraphics[width=4.2cm]{Prob_B_piby20_Vo_piby2_kx_piby10.eps}}
\subfigure[]{\includegraphics[width=4.2cm]{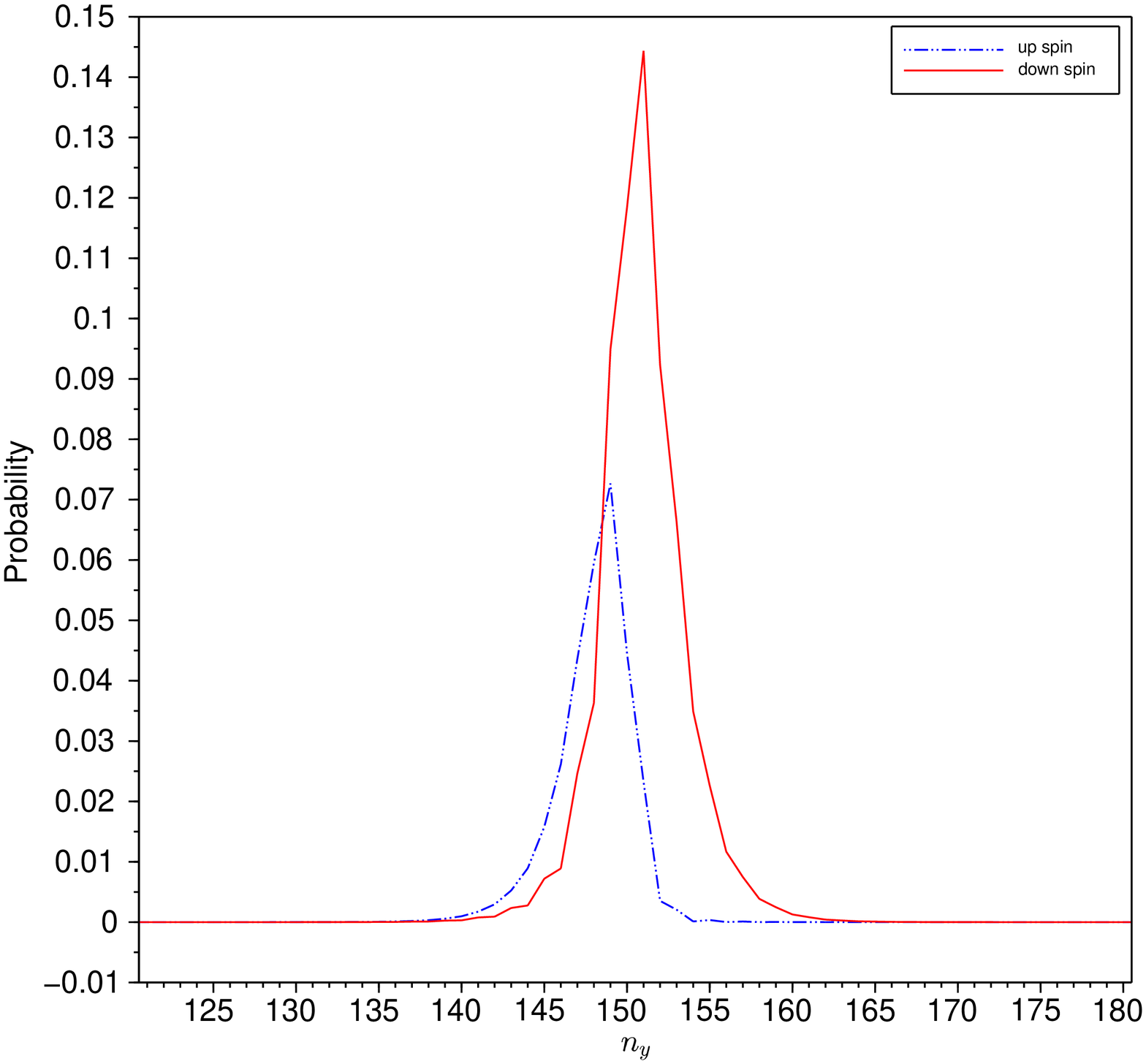}}
\end{center}
\caption{Probabilities of barrier states for $V_b=\pi/2$ and (a) $B=0$, $k_x 
= -\pi/10$, (b) $B=\pi/20$, $k_x = -\pi/10$, (c) $B=0$, $k_x = 0$, (d) $B =
\pi/20$, $k_x = 0$, (e) $B=0$, $k_x = \pi/10$, (f) $B=\pi/20$, $k_x = \pi/10$.
The probabilities of spin-$\ua$ and $\da$ are shown in blue (dashed line) and 
red (solid line) respectively.} \label{fig:prob} \end{figure}

The states produced by the potential barrier have probabilities which decay 
exponentially as we go away from the centre of the Gaussian. The probabilities
of spin-$\ua$ and $\da$ are given by $|\al_{n_y}|^2$ and $|\be_{n_y}|^2$ 
respectively. For $B=0$, these are shown in Figs.~\ref{fig:prob} (a), (c) 
and (e) for states with $k_x =-\pi/ 10$, 0 and $\pi/10$ respectively. We see 
that for $k_x =0$, the probability is spread over the entire range of $n_y$.
(The probability looks like a band because it oscillates between 0 and $0.007$ 
with period 2 in $n_y$). Hence this state is not localized; this will be 
studied further below using the decay length. 

Note that all the bound states shown in Figs.~\ref{fig:prob}, namely, (a), 
(b), (d), (e) and (f), have probability profiles which are quite smooth; they 
have small oscillations with period 2 in $n_y$ due to fermion doubling, but 
these are not visible in the figures. This implies that choosing a smooth
potential profile has enabled us to essentially bypass the fermion doubling 
problem.

For $B=\pi/20$, Figs.~\ref{fig:prob} (b), (d) and (f) show the probabilities 
for $k_x = -\pi/10$, 0 and $\pi/10$ respectively. In this case, the $k_x=0$ 
state is also localized. In Figs.~\ref{fig:prob} (b), (d) and (f), the
spin-$\da$ probabilities are larger than the spin-$\ua$ probabilities because 
of the presence of a magnetic field $B > 0$. In all the plots in
Figs.~\ref{fig:prob} we observe that the probabilities of spin-$\ua$ and $\da$
get reflected about the centre of the Gaussian when we change $k_x \to - k_x$;
this is a consequence of both the symmetries $\cal P$ and $\cal R$.

\begin{figure}[htb]
\begin{center}
%\subfigure[]{\includegraphics[width=4.27cm]{B_0_Vo_piby2_up_decay_length_a_6.0043501.eps}}
\subfigure[]{\includegraphics[width=4.27cm]{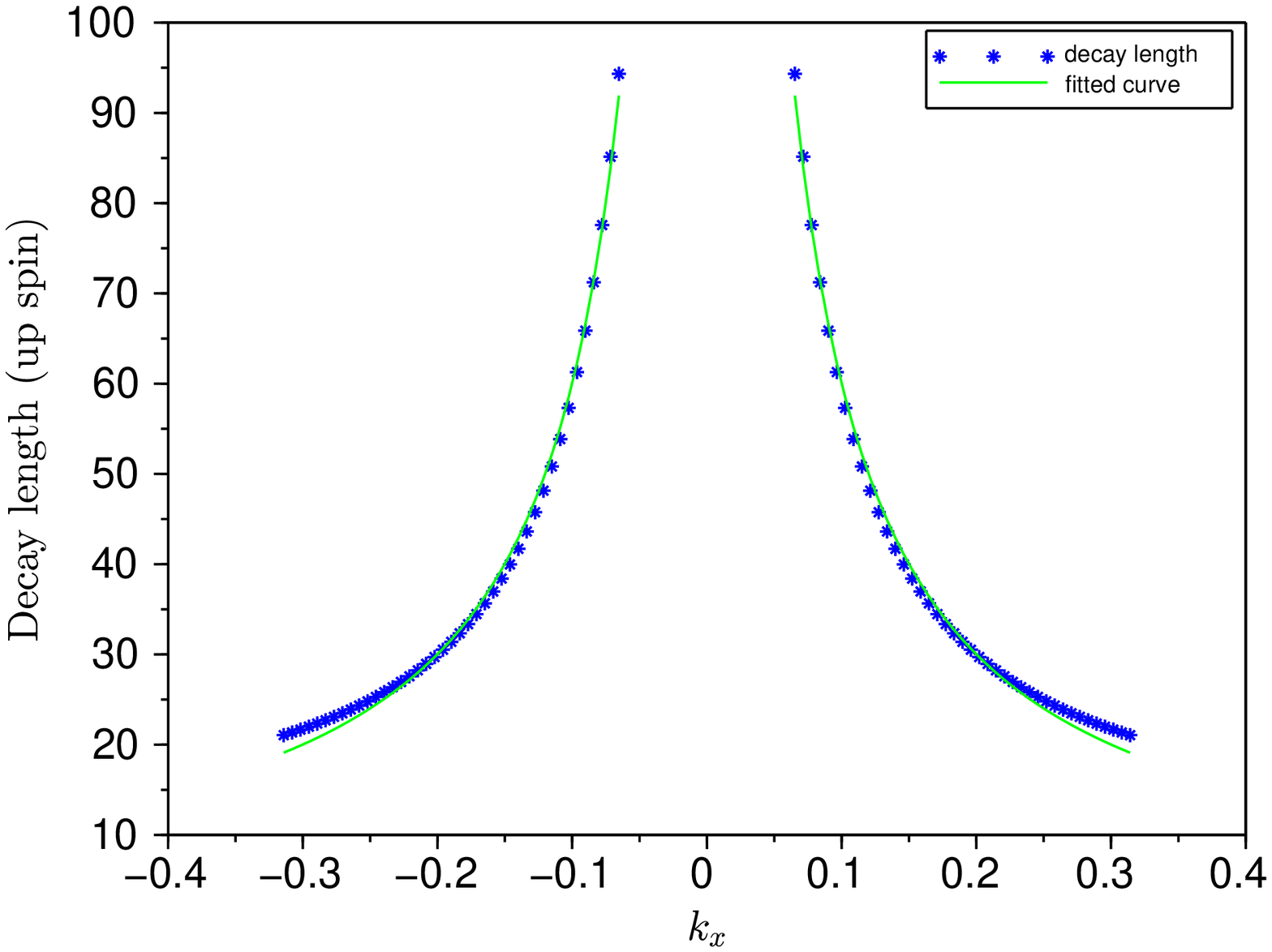}}
%\subfigure[]{\includegraphics[width=4.27cm]{B_0_Vo_piby2_dn_decay_length_a_5.9849056.eps}}
\subfigure[]{\includegraphics[width=4.27cm]{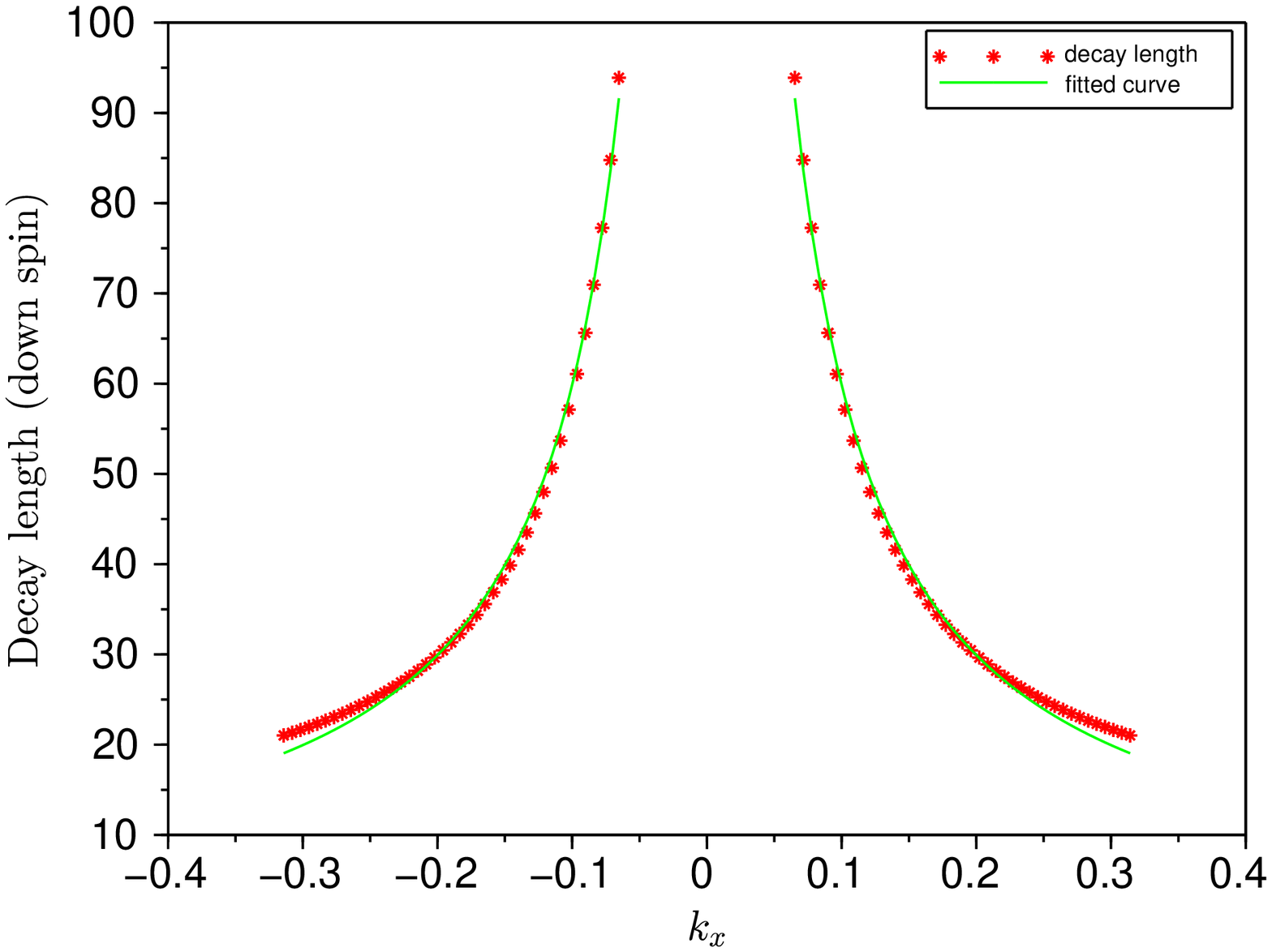}}
\end{center}
\caption{Estimates of decay length $\xi$ (shown by blue circles) from inverse 
participation ratio of barrier states for $V_b=\pi/2$ and $B=0$ for (a) 
spin-$\ua$ and (b) spin-$\da$. A least square fit of the form $\xi = c/|k_x|$ 
(shown by green lines) gives $c=6.00$ for spin-$\ua$ and $5.98$ for 
spin-$\da$.} \label{fig:decay} \end{figure}

{\bf Decay length:} Since the probabilities in Figs.~\ref{fig:prob} decay
rather rapidly (within a few lattice spacings), it is difficult to estimate 
the decay lengths accurately from these probabilities. The decay length can 
be estimated more easily from the IPR as follows. Since the states
decay in only direction, the probability $|\psi^2|$ will be proportional
to $e^{-|n|/\xi}/\xi$ (where $n$ denotes the deviation of $n_y$ from the 
centre of the Gaussian), and the IPR will be proportional to $1/\xi$.
We therefore simply define the decay length $\xi$ to be the inverse of the 
IPR and plot the resultant values of $\xi$ versus the momentum $k_x$. We 
find that $\xi$ is proportional to $1/|k_x|$, the constant of proportionality
being almost the same for the probabilities of the spin-$\ua$ and $\da$
components. This is shown in Fig.~\ref{fig:decay} for $V_b=\pi/2$ and $B=0$.
We observe that the decay length diverges as $k_x \to 0$, implying that there 
is no bound state at $k_x=0$. Thus the spectrum of bound states does not 
contain the point $k_x = 0$. The situation is quite different when a magnetic
field is present; then the decay length is finite for all values of $k_x$ 
and there is a bound state even when $k_x = 0$.

\begin{figure}[htb]
\begin{center}
\includegraphics[width=5.7cm]{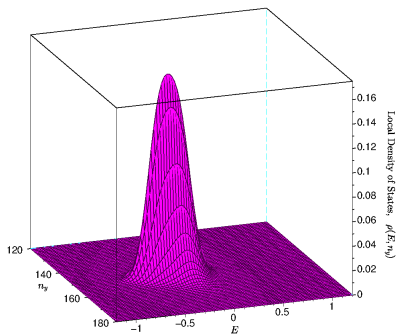}
\end{center}
\caption{Local density of states due to bound states produced by a potential 
barrier with $V_b = \pi/2$ and a magnetic field $B = \pi/20$. The 
$\de$-functions in energy have been replaced by Gaussians with width $0.1$.} 
\label{fig:ldos} \end{figure}

{\bf Local density of states:} It is useful to look at the local density 
of states produced by the potential barrier. For the case $V_b = \pi/2$ and 
$B=\pi/20$ where there is an almost flat band (Fig.~\ref{fig:disp} (d)),
the local density of states produced by the bound states lying in the 
range $-\pi/5 < k_x < \pi/5$ is defined to be
\beq \rho (E,n_y) ~=~ \int_{-\pi/5}^{\pi/5} ~\frac{dk_x}{2\pi} ~\de 
(E - E_{k_x}) ~|\psi (k_x;n_y)|^2. \label{ldos} \eeq
This is shown in Fig.~\ref{fig:ldos} where we have smoothened the 
$\de$-functions in Eq.~\eqref{ldos} by replacing them by Gaussians with width 
$0.1$. (We have approximated the integral in Eq.~\eqref{ldos} by taking a large
number $N$ of equally spaced points in $k_x$ from $-\pi/5$ to $\pi/5$, adding 
up the contributions from all those points, and dividing by $5N$). As expected
from Fig.~\ref{fig:disp} (d), the local density of states is peaked at an 
energy of about $-0.2$ and at $n_y = 151$ where the barrier is located.

\begin{figure}[htb]
\begin{center}
%\subfigure[]{\includegraphics[width=4.27cm]{Spin_B_0_Vo_piby2.eps}}
\subfigure[]{\includegraphics[width=4.27cm]{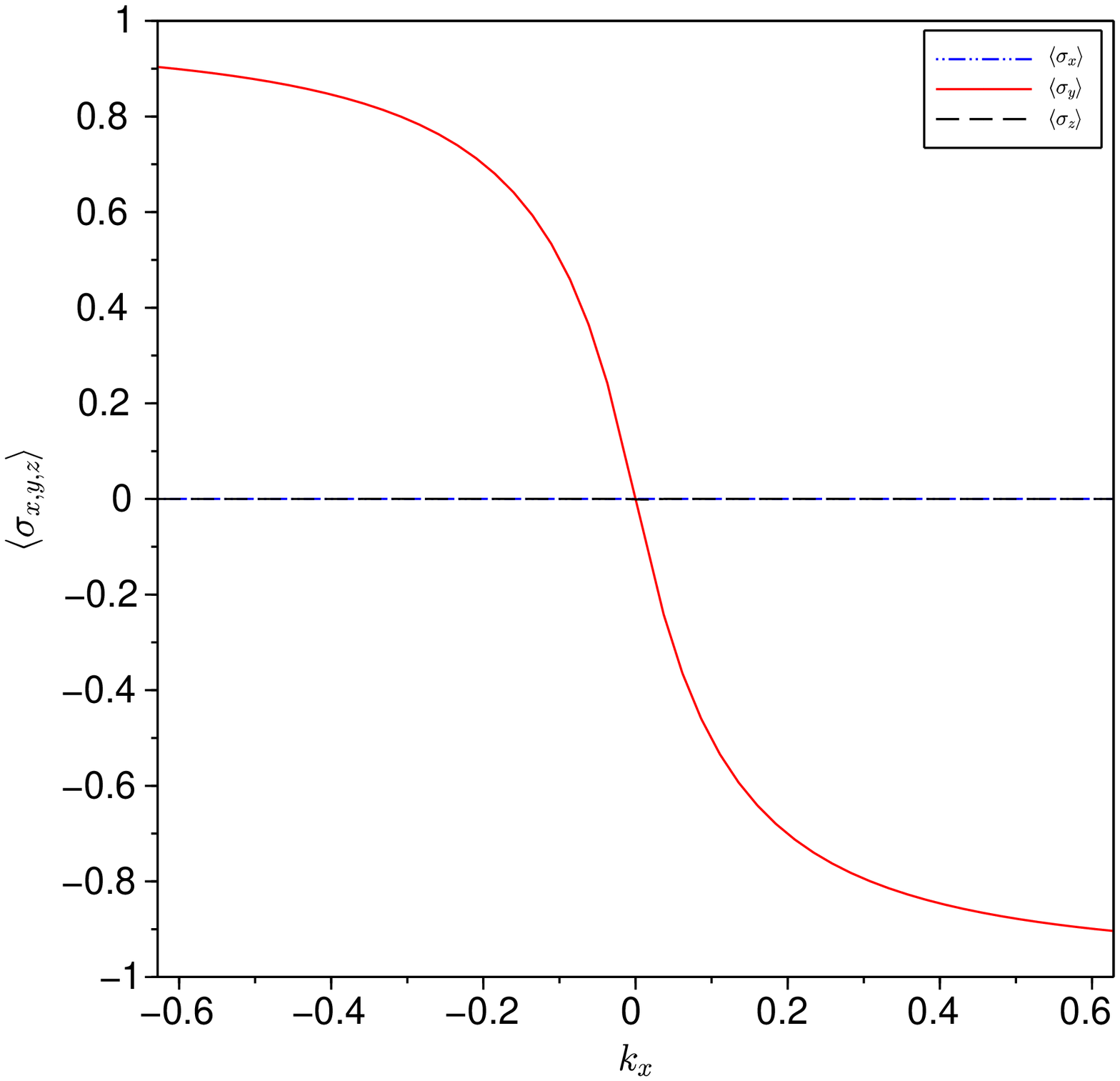}}
%\subfigure[]{\includegraphics[width=4.27cm]{Spin_B_piby20_Vo_piby2.eps}}
\subfigure[]{\includegraphics[width=4.27cm]{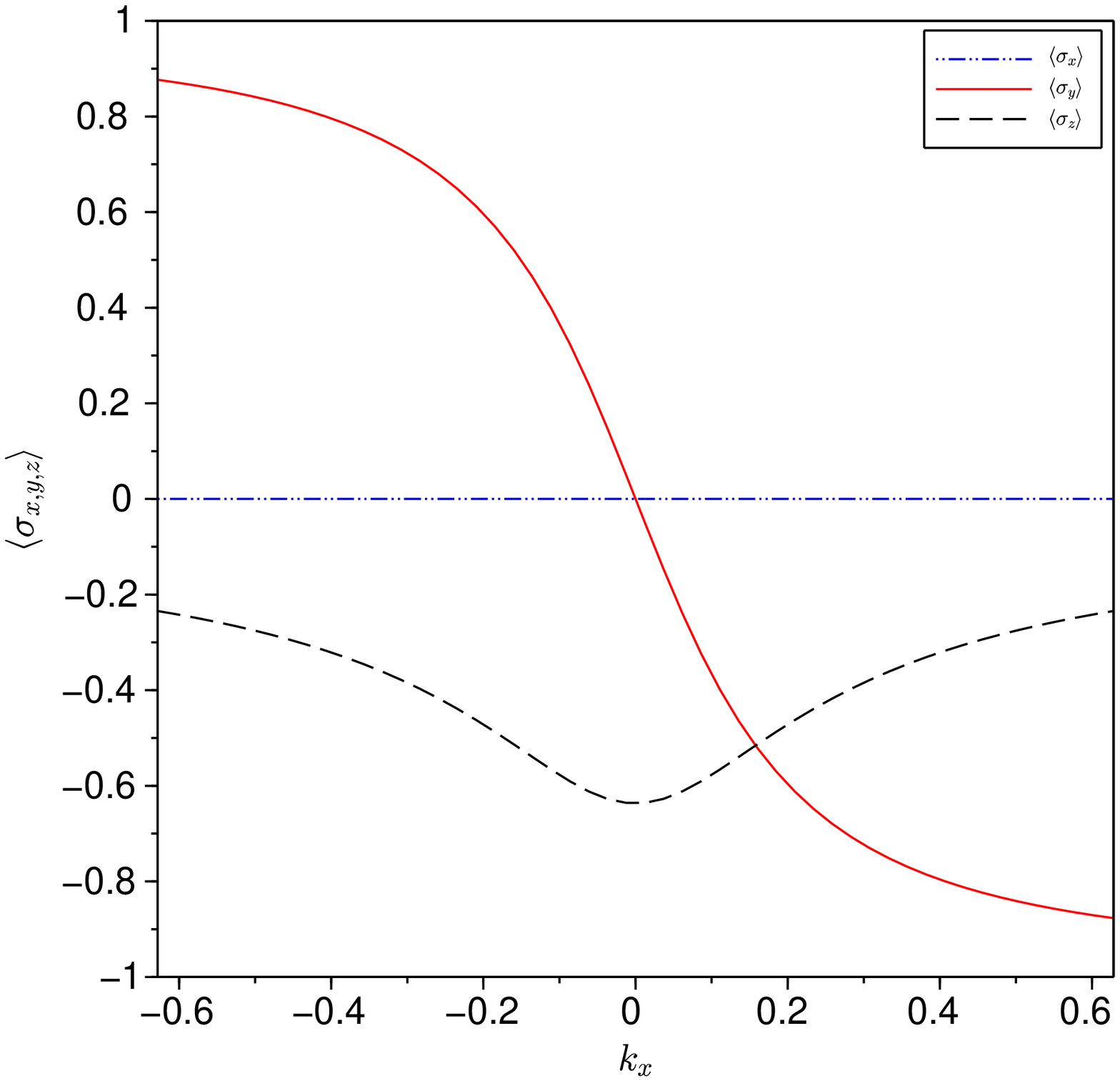}}
\end{center}
\caption{Spin expectation values of barrier states for (a) $V_b=\pi/2$, $B=0$,
and (b) $V_b=\pi/2$, $B=\pi/20$. The $\la \si^x \ra$, $\la \si^y \ra$ and
$\la \si^z \ra$ are shown by blue dot-dash, red solid and black dashed lines 
respectively. For $B=0$, $\la \si^x \ra = \la \si^z \ra = 0$ coincide.} 
\label{fig:spin} \end{figure}

{\bf Spin:} It is interesting to look at the expectation values of 
the different components 
of the spin as a function of $k_x$. This is shown in Fig.~\ref{fig:spin} for
$V_b=\pi/2$ and (a) $B=0$ and (b) $B=\pi/20$. These figures show certain 
symmetries which can be understood as follows. The symmetry $\cal P$ under 
which $\psi (x,y) \to \psi^* (x,-y)$ implies that $\la \si^y \ra$ will change 
sign but $\la \si^x \ra$ and $\la \si^z \ra$ will remain the same if we change
$k_x \to - k_x$. The symmetry $\cal R$ under which $\psi (x,y) \to \si^z \psi 
(-x,-y)$ implies that $\la \si^x \ra$ and $\la \si^y \ra$ will change sign but
$\la \si^z \ra$ will remain the same under $k_x \to - k_x$. Combining
these results, we see that $\la \si^x \ra$ must be equal to zero for each
value of $k_x$ for any value of $B$; this agrees with Fig.~\ref{fig:spin}. 
Finally, if $B=0$, time-reversal symmetry leads to all three spin 
expectation values changing sign under $k_x \to - k_x$. Combined with 
the symmetries $\cal P$ or $\cal R$, this means that $\la \si^z \ra$ must
equal zero for each value of $k_x$.

To summarize this section, applying a combination of a translation invariant 
potential barrier and a magnetic field to Dirac electrons can produce an
one-dimensional system which can be thought of as a one-band
quantum wire. The dispersion of this system is unusual
in that the energy is an even function of $k_x$, unlike
chiral systems where the energy is an odd function such as $E = v k_x$.
The dispersion can be controlled by tuning the barrier strength; in 
particular, the dispersion can be made almost flat. The wave functions of
these states decay exponentially away from the barrier; the decay length
increases as $|k_x|$ decreases but remains finite at $k_x=0$ if a magnetic
field is present. The expectation values of the spin components also vary
with $k_x$. We note that all these results are in qualitative agreement with 
those obtained analytically in Ref.~\onlinecite{deb1} for the case of a 
$\de$-function potential barrier.

\section{Numerical results in two dimensions}
%\section{Lattice model in two dimensions}

In this section, we will present our results for three cases where the 
potential $V(n_x,n_y)$ does not have translational symmetry along any 
direction. We will therefore use Eqs.~(\ref{eq:lat1}-\ref{eq:lat2}) 
to find the spectrum for a two-dimensional system. 

Apart from the symmetries $\cal T$ (if $B=0$), $\cal P$ (if $V(n_x,-n_y)=
V(n_x,n_y)$) and $\cal R$ (if $V(-n_x,-n_y)= V(n_x,n_y)$) which the continuum 
Hamiltonian has, Eqs.~(\ref{eq:lat1}-\ref{eq:lat2}) has two other symmetries 
which are peculiar to the lattice model. Eqs.~(\ref{eq:lat1}-\ref{eq:lat2}) 
remain invariant under a transformation ${\cal A}_x$ which takes 
$\psi_{n_x,n_y} \to (-1)^{n_x} \si^z \psi^*_{n_x,n_y}$, and a transformation 
${\cal A}_y$ which takes $\psi_{n_x,n_y} \to (-1)^{n_y} \psi^*_{n_x,n_y}$. 
Combining ${\cal A}_x$ and ${\cal A}_y$, we find a symmetry under the 
transformation ${\cal A}_{xy}$ which takes $\psi_{n_x,n_y} \to (-1)^{n_x+ n_y}
\si^z \psi_{n_x,n_y}$. This implies that the eigenstates of the Hamiltonian 
can be chosen to satisfy either ${\cal A}_{xy} \psi = \psi$ or 
${\cal A}_{xy} \psi = -\psi$. If ${\cal A}_{xy} \psi = \psi$, the spin-$\ua$
(spin-$\da$) component must be zero if $n_x + n_y$ is odd (even), and the
situation is reversed if ${\cal A}_{xy} \psi = - \psi$. Thus imposing the 
constraint ${\cal A}_{xy} \psi = \psi$ (or $= -\psi$) would eliminate half 
the components of $\psi_{n_x,n_y}$. This is equivalent to the Kogut-Susskind 
prescription for reducing the fermion doubling problem; the reduction is by 
a factor of two in this system~\cite{kogut}. 

However, we will {\it not} impose constraints of the form ${\cal A}_{xy} \psi 
= \pm \psi$ when doing our numerical conditions since this would lead to wave 
functions which have large oscillations of period 2 in $n_x$ and $n_y$. The 
various wave functions that we have found numerically and have discussed below
are all quite smooth and have only small oscillations of period 2. Once again,
this is because we have chosen all the potentials to have very small Fourier 
components near $k_x$ or $k_y$ equal to $\pi$.

\begin{figure}[htb]
\begin{center}
%\subfigure[]{\includegraphics[width=5cm]{Surface_Potential_B_0.1570796_Vimp_15.707963.eps}} \\
\subfigure[]{\includegraphics[width=5cm]{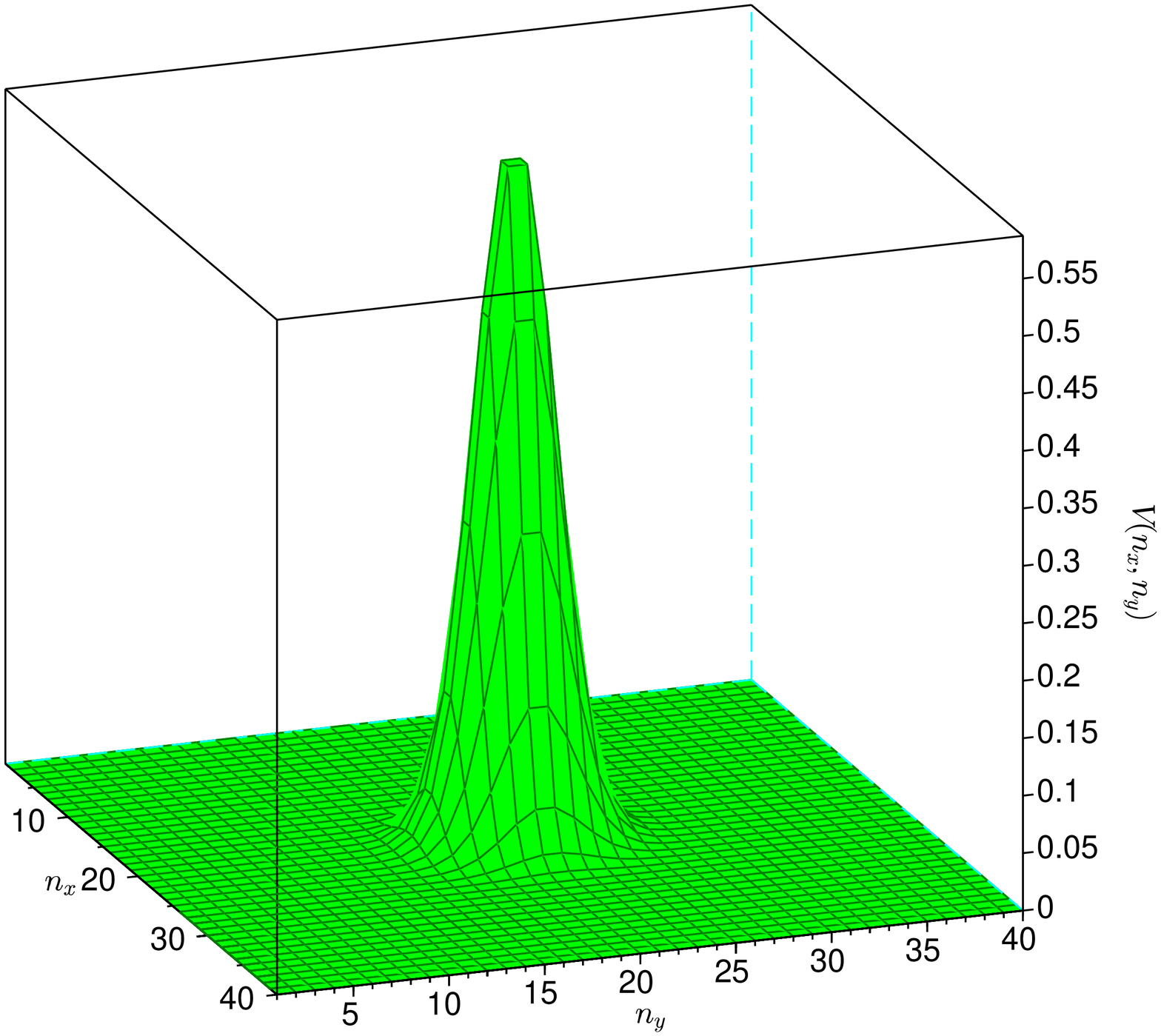}} \\
%\subfigure[]{\includegraphics[width=4.25cm]{Prob_up_B_piby20_Vimp_5pi_sigma_2_state_1600.eps}} 
\subfigure[]{\includegraphics[width=4.25cm]{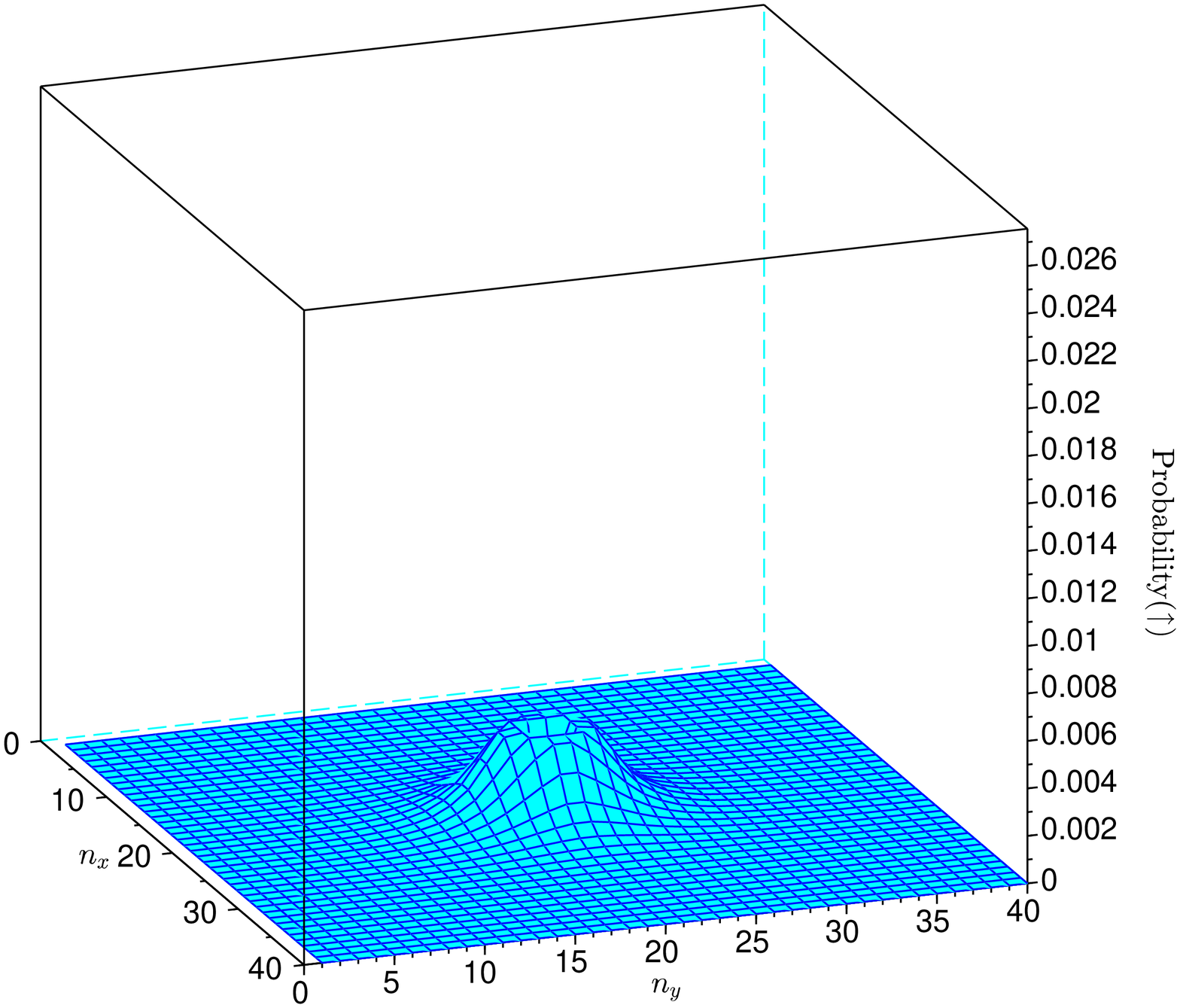}} 
%\subfigure[]{\includegraphics[width=4.25cm]{Prob_dn_B_piby20_Vimp_5pi_sigma_2_state_1600.eps}}
\subfigure[]{\includegraphics[width=4.25cm]{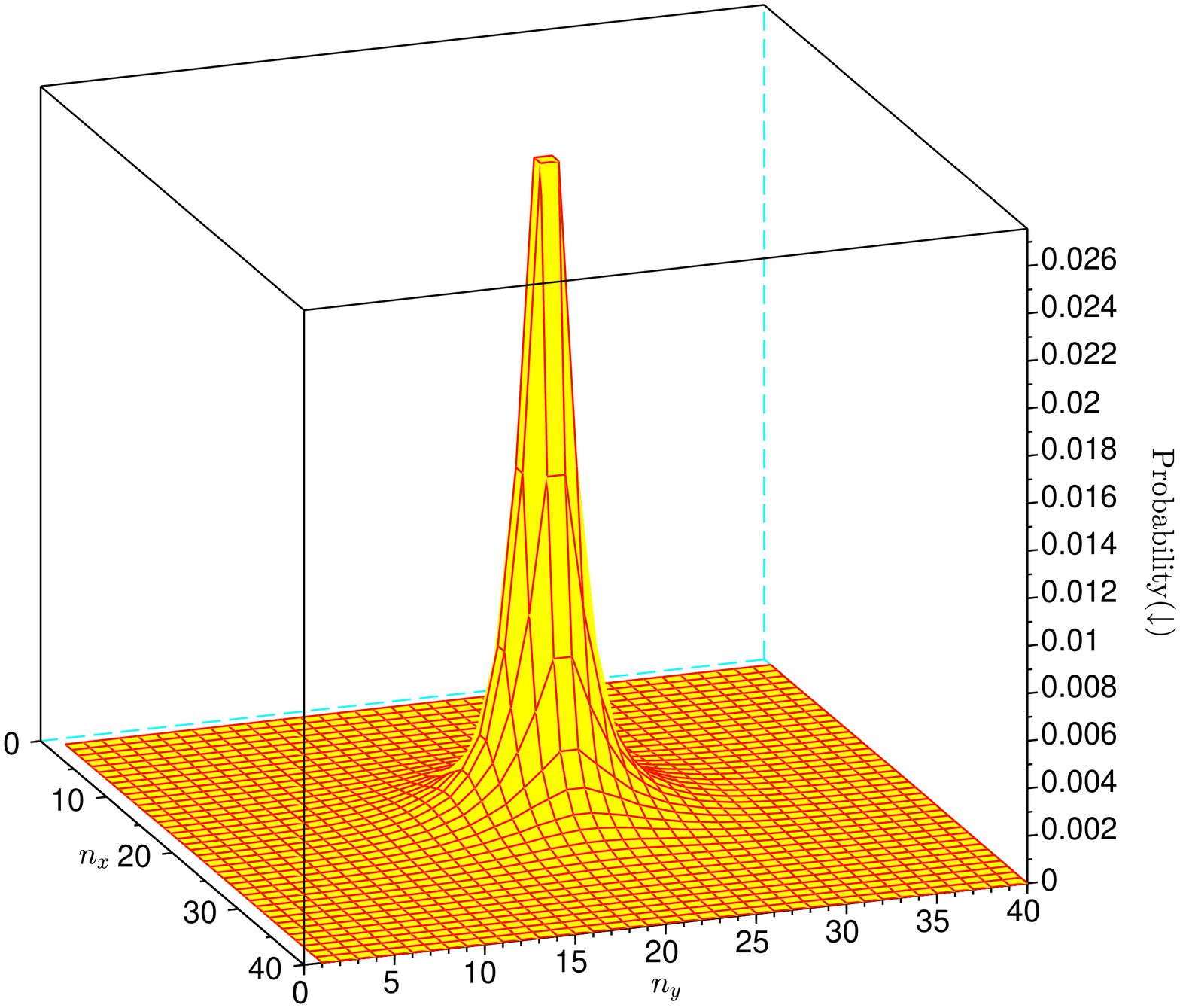}}
\end{center}
\caption{(a) Surface plot of an impurity potential which is a Gaussian with 
width 2 in both directions and $V_i = 5\pi$, as described in 
Eq.~\eqref{eq:vimp}. 
Surface plots of probabilities of (b) spin-$\ua$ and (c) spin-$\da$ for the 
bound state, for $B = \pi/20$ and a system with $40 \times 40$ sites. The 
spin-$\da$ probability is much larger than the spin-$\ua$ probability; this 
is because the magnetic field points in the $+\hat z$ direction.} 
\label{fig:impurity} \end{figure}

{\bf Impurity potential:} We first consider the effect of a localized 
potential which may arise from a
non-magnetic impurity; by localized we mean that the potential rapidly goes 
to zero outside some region. In particular, we will consider a Gaussian form
\beq V_{imp;n_x,n_y} ~=~ \frac{V_i}{2\pi \si^2} ~e^{-[(n_x - n_0)^2 + 
(n_y - n_0)^2)/(2\si^2)}. \label{eq:vimp} \eeq
where $\si =2$ and $V_i = 5 \pi$. In the absence of a magnetic field, we 
find numerically that this potential does not produce a bound state.
However, when we turn on a magnetic field (we take $B=\pi/20$), we find that
a localized bound state can appear as shown in Fig.~\ref{fig:impurity}.
(The inverse participation ratio is particularly large for states which are 
localized in both directions and is therefore very useful for numerically
finding such states). 
We thus see that while a potential alone does not localize a Dirac electron
(since an electron can Klein tunnel through a potential), a potential along 
with a magnetic field can localize an electron. Qualitatively, this is
because a magnetic field produces a gap; if a localized potential can produce 
a state lying in that gap, the wave function of that state will decay
exponentially as one goes far away from the potential thereby producing a
localized state. This suggests that one can construct a quantum dot hosting
one or more states by applying a localized potential and a magnetic field
to a system of Dirac electrons.

\begin{figure}[htb]
\begin{center}
%\subfigure[]{\includegraphics[width=6cm]{Barrier_impurity_V_piby2_Vimp_pi_sigma_2.eps}} \\
\subfigure[]{\includegraphics[width=6cm]{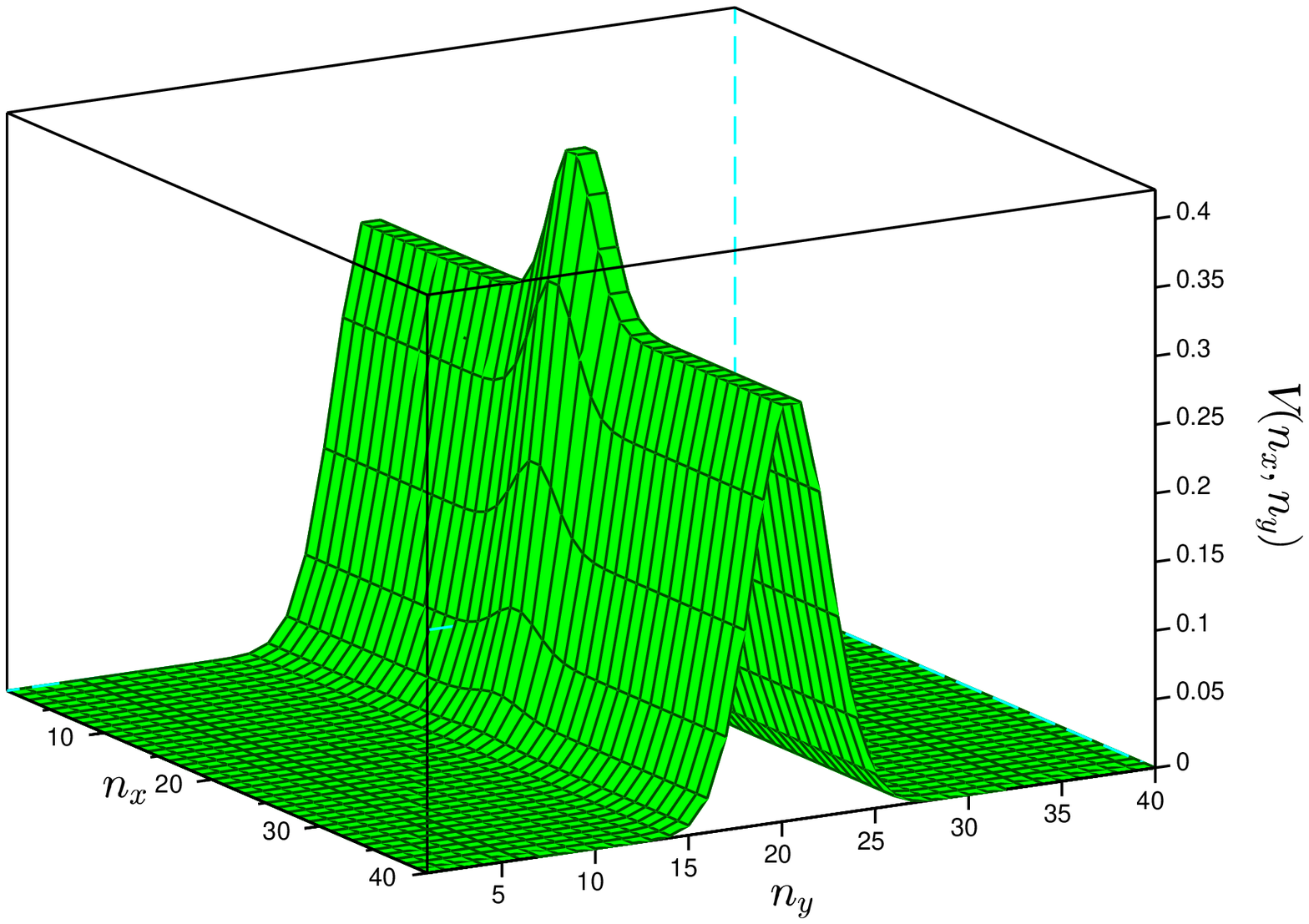}} \\
%\subfigure[]{\includegraphics[width=4.25cm]{Prob_up_B_piby20_Vo_piby2_Vimp_pi_sigma_2_state_1600.eps}}
\subfigure[]{\includegraphics[width=4.25cm]{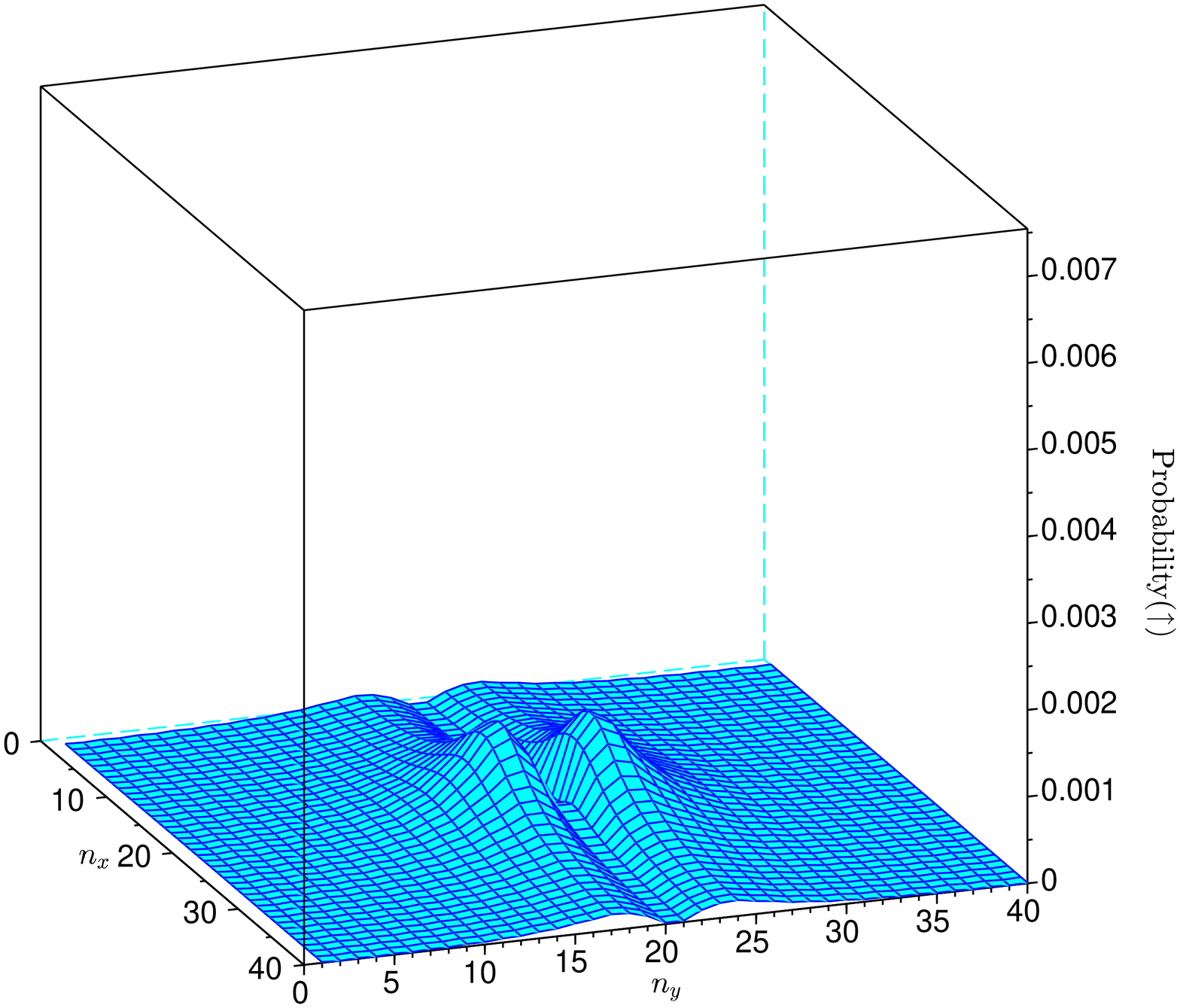}}
%\subfigure[]{\includegraphics[width=4.25cm]{Prob_dn_B_piby20_Vo_piby2_Vimp_pi_sigma_2_state_1600.eps}}
\subfigure[]{\includegraphics[width=4.25cm]{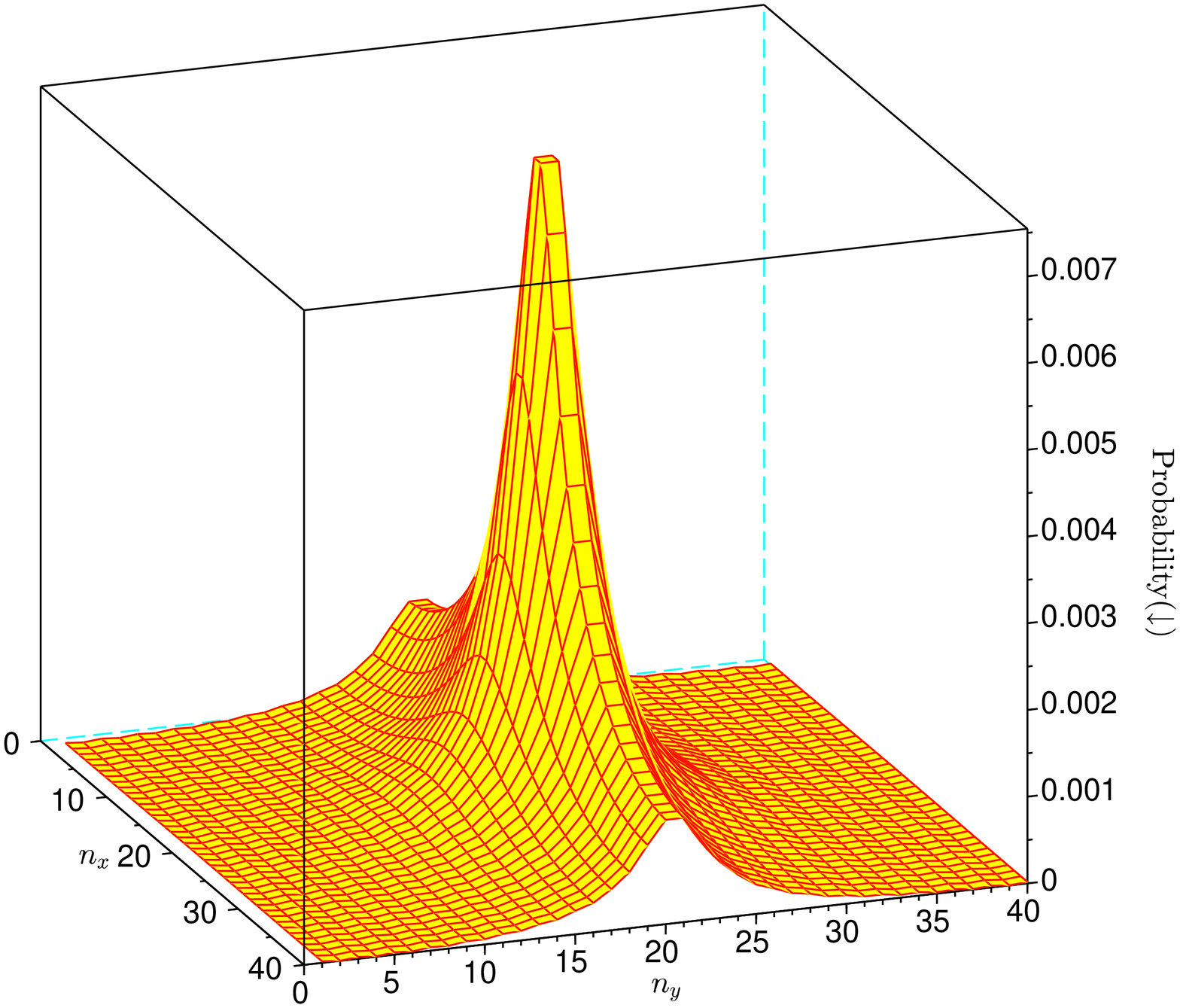}}
\end{center}
\caption{(a) Surface plot of a potential which is a combination of a barrier 
which is a Gaussian with width 2 and $V_b = \pi/2$ in the transverse direction
and an impurity potential which is a Gaussian with width 2 in both directions 
and $V_i = \pi$. Surface plots of probabilities of (b) spin-$\ua$ and (c) 
spin-$\da$ for the bound state, for $B = \pi/20$ and a system with $40 \times 
40$ sites.} \label{fig:barrierplusimpurity} \end{figure}

{\bf Potential barrier and impurity:}
Next we consider when both a barrier and an impurity are 
present (both are assumed to be spin independent). In the absence of a 
magnetic field, the states propagating as plane waves along the barrier 
are not expected to scatter from the impurity due to time-reversal symmetry.
For a weak impurity, this can be shown using first order perturbation theory.
For an elastic (i.e., energy conserving scattering), we only need to consider
scattering between two plane wave states with equal and opposite momenta 
$\pm k_x$. Let us denote the corresponding wave functions as 
$\psi_{k_x;n_x,n_y}$ and $\psi_{-k_x} (n_x,n_y)$. Then the Born approximation 
in one dimension~\cite{agarwal} shows that the reflection amplitude produced 
by an impurity $V_{imp;n_x,n_y}$ is given by
\beq r_{k_x} ~=~ - ~\frac{i}{dE/dk_x} ~\la \psi_{-k_x} | V_{imp} | \psi_{k_x} 
\ra. \label{refl} \eeq
In the absence of a magnetic field, $\psi_\pm$ are related by time-reversal
transformation: 
\beq \psi_{-k_x;n_x,n_y} ~=~\si^y \psi^*_{k_x;n_x,n_y}. \label{timrev} \eeq
Hence the matrix element in Eq.~\eqref{refl} is equal to 
\bea && \la \psi_{-k_x} | V_{imp} | \psi_{k_x} \ra \non \\
&& = \sum_{n_x,n_y} V_{imp;n_x,n_y} \psi^T_{k_x;n_x,n_y} \si^y 
\psi_{k_x;n_x,n_y}. \eea
This vanishes for any form of $V_{imp}$ because the antisymmetry of $\si^y$ 
implies that $\psi^T_{k_x;n_x,n_y} \si^y \psi_{k_x;n_x,n_y} = 0$. Thus 
the barrier states are immune to scattering by weak impurities if no magnetic 
field is present. This also implies that an impurity cannot produce a bound 
state. This is because bound states in one dimension occur at the complex 
values of $k_x$ where $r_{k_x}$ has a pole (when $r_{k_x}$ is analytically
continued away from the real axis); if $r_{k_x} = 0$ for all $k_x$, its
analytical continuation will also be zero and it will have no pole in
the complex plane.

These arguments break down when a magnetic field is present because 
$\psi_{\pm k_x}$ will no longer be related to each other by 
Eq.~\eqref{timrev}, and $\la \psi_{-k_x} | V_{imp} | \psi_{k_x} \ra$ will not 
be equal to zero in general; hence the reflection amplitude in Eq.~\eqref{refl}
will no longer vanish. In addition, a bound state becomes possible. This is
illustrated in Fig.~\ref{fig:barrierplusimpurity} where the spin-$\ua$ and 
$\da$ probabilities are shown for a bound state which appears when there is a 
potential barrier and an impurity with the forms given in Eqs.~\eqref{eq:barr}
and \eqref{eq:vimp} with $V_b = \pi/2$ and $V_i = \pi$ respectively, and a 
magnetic field $B=\pi/20$ is also present. The spin-$\da$ probability again 
turns out to be much larger than the spin-$\ua$ probability because the 
magnetic field points in the $+\hat z$ direction.

Assuming that a magnetic field is present and $\la \psi_{-k_x} | V_{imp} | 
\psi_{k_x} \ra \ne 0$, Eq.~\eqref{refl} implies that the 
reflection amplitude is larger if the group velocity $dE/dk_x$ is smaller. 
This means that if the barrier strength is tuned to produce an almost flat 
band, even a small impurity potential will lead to a large backscattering.

We find numerically that for a given value of the magnetic
field, the strength of the impurity potential which is required to produce a 
state bound to it is smaller when a potential barrier is present compared to 
the case when a potential barrier is not present. This is why we set $V_i =
\pi$ in Fig.~\ref{fig:barrierplusimpurity} but $V_i = 5 \pi$ in 
Fig.~\ref{fig:impurity}. A qualitative reason for this is that a potential 
barrier already creates edge states which are localized in one direction 
(perpendicular to the barrier); then an impurity potential only has to 
localize such a state in the other direction (along the barrier). Without a 
potential barrier, the impurity potential by itself has to localize a bound 
state in two directions.

\begin{figure}[htb]
\begin{center}
\includegraphics[width=6cm]{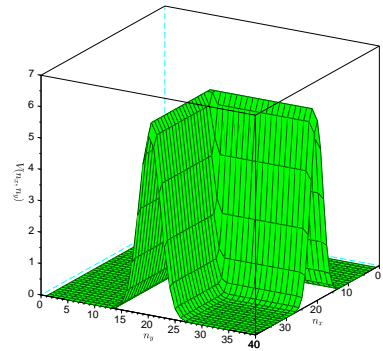}
\end{center}
\caption{Surface plot of an ``L"-shaped potential. The potential has a width of
2 and a maximum value of $5\pi/(2\sqrt{2\pi})$ along the spine of the ``L".} 
\label{fig:Lshapepotential} \end{figure}

\begin{figure}[htb]
\begin{center}
%\subfigure[]{\includegraphics[width=4.25cm]{Prob_up_B_piby20_Vo_5pi_sigma_2_state_1023.eps}}
\subfigure[]{\includegraphics[width=4.25cm]{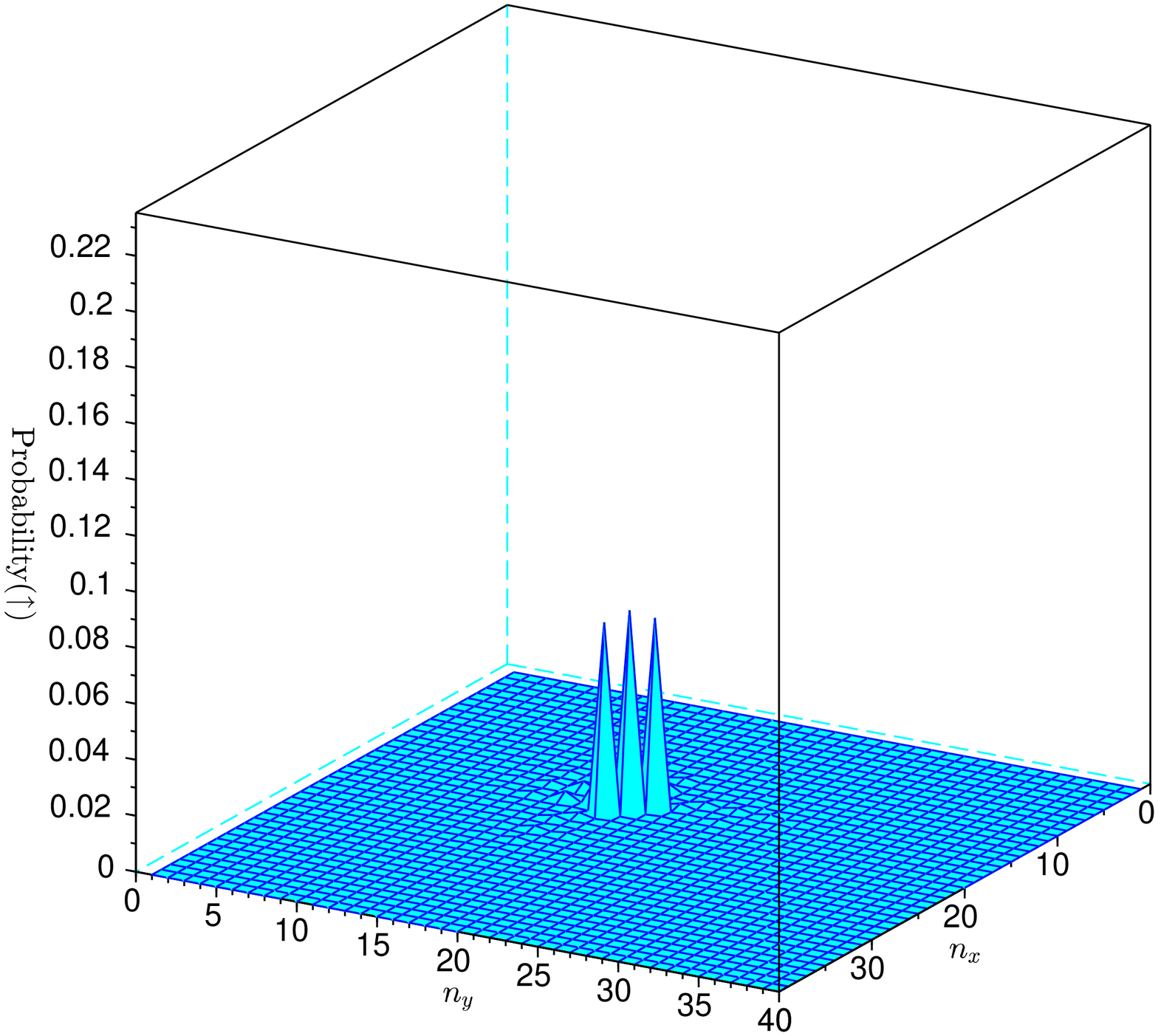}}
%\subfigure[]{\includegraphics[width=4.25cm]{Prob_dn_B_piby20_Vo_5pi_sigma_2_state_1023.eps}}
\subfigure[]{\includegraphics[width=4.25cm]{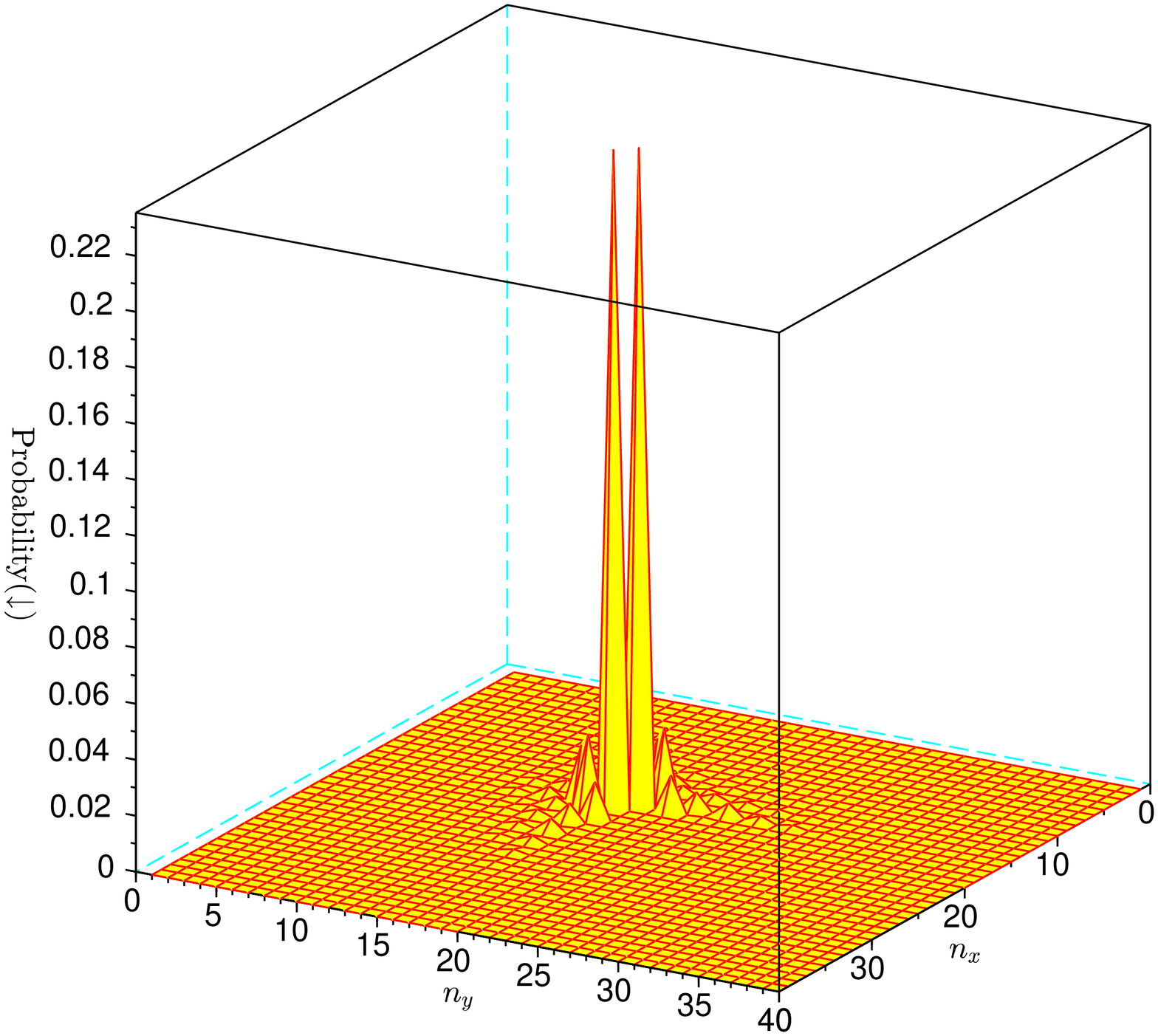}}
\end{center}
\caption{Surface plots of probabilities of (a) spin-$\ua$ and (b) spin-$\da$ 
for the bound state localized at the corner of an ``L"-shaped potential, for 
$B = \pi/20$ and a system with $40 \times 40$ sites.} 
\label{fig:Lshapecorner} \end{figure}

\begin{figure}[htb]
\begin{center}
%\subfigure[]{\includegraphics[width=4.25cm]{Prob_up_B_piby20_Vo_5pi_sigma_2_combo_iplus1.eps}}
\subfigure[]{\includegraphics[width=4.25cm]{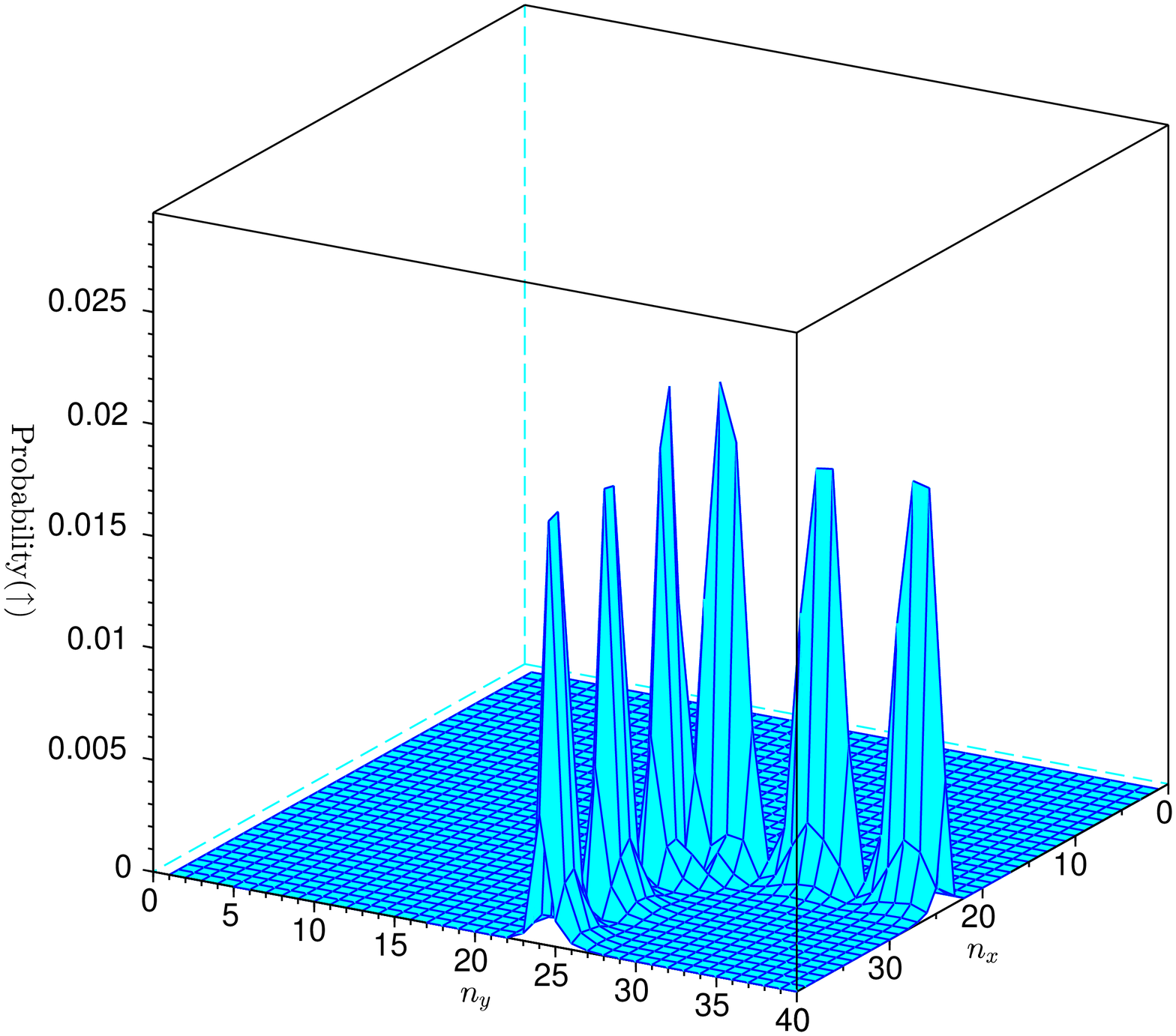}}
%\subfigure[]{\includegraphics[width=4.25cm]{Prob_dn_B_piby20_Vo_5pi_sigma_2_combo_iplus1.eps}}
\subfigure[]{\includegraphics[width=4.25cm]{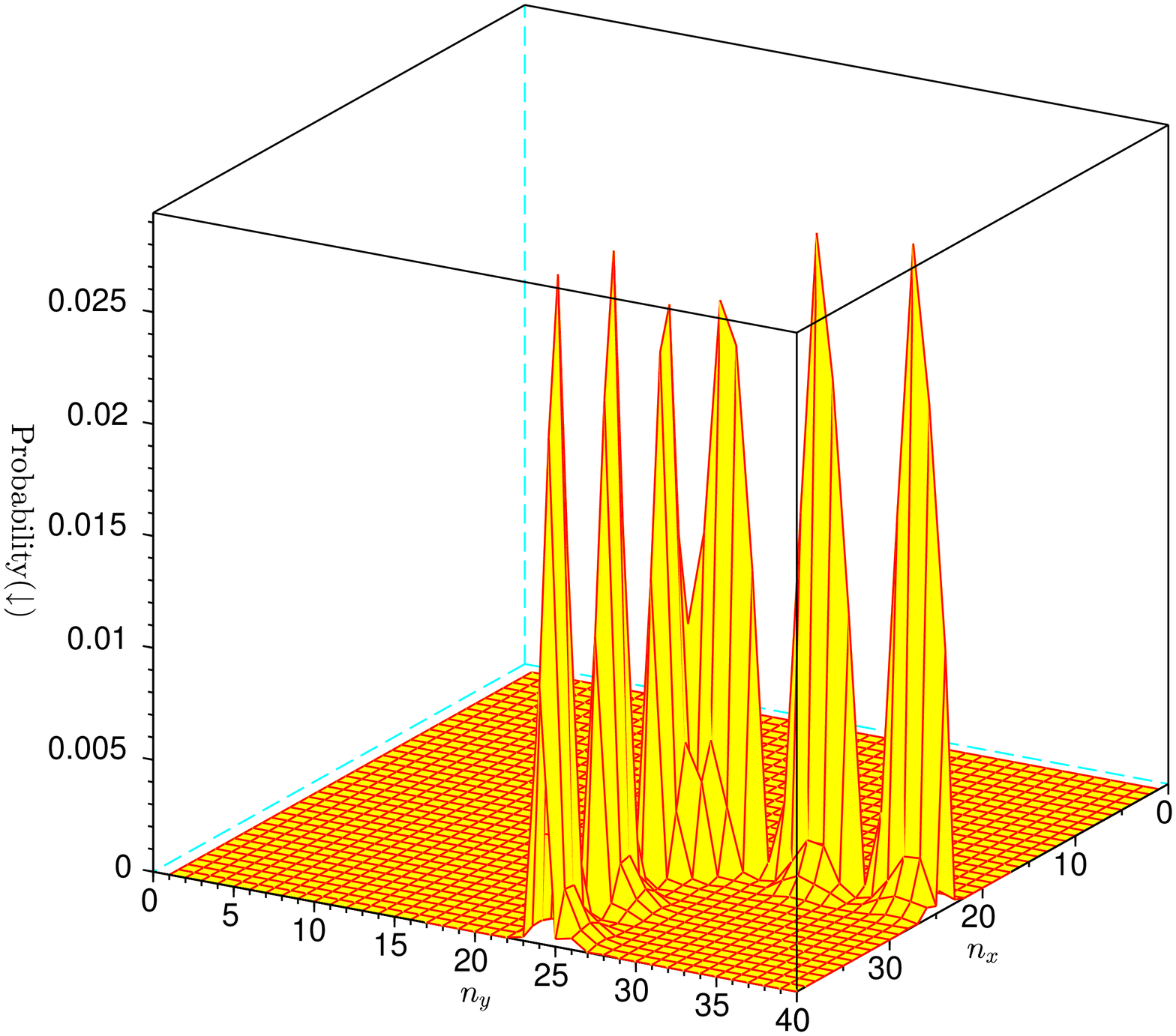}}
\end{center}
\caption{Surface plots of probabilities of (a) spin-$\ua$ and (b) spin-$\da$ 
for a state which is extended along the arms of an ``L"-shaped potential, for 
$B = \pi/20$ and a system with $40 \times 40$ sites.} \label{fig:Lshapearm} 
\end{figure}

{\bf ``L"-shaped potential:}
Finally, we give an example to show that one can create quasi-one-dimensional
systems with bends which can host either localized or extended states of 
electrons. Fig.~\ref{fig:Lshapepotential} shows an ``L"-shaped potential 
barrier consisting of two semi-infinite arms in perpendicular directions; each
arm has the form given in Eq.~\eqref{eq:barr} with $\si=2$ and $V_b = 5 \pi$. 
In general we find two kinds of states, one localized at the 
the corner of the ``L" and the other running along the arms. An example of
a bound state localized at the corner is shown in Fig.~\ref{fig:Lshapecorner}.
Fig.~\ref{fig:Lshapearm} shows an extended wave function which 
runs along the arms of the potential. (The spin probabilities for this state
are in the form of standing waves because of reflections from the edges of
the system where we have used open boundary conditions. The wave function 
would have been a plane wave instead of a standing wave if the system was
infinitely large). The ratio of the spin-$\da$ probability to the spin-$\ua$ 
probability is much larger for the state in Fig.~\ref{fig:Lshapecorner} 
compared to the state in Fig.~\ref{fig:Lshapearm}. This is because the 
magnitudes of the momenta $k_x$ and $k_y$ (one or both of which must be 
complex for a state which is localized in one or both directions) turns out 
to be much smaller than the magnetic field $B$ for the localized state; hence 
the wave function is dominated by the magnetic field and therefore has a large
component in the direction opposite to it. For the extended state, however, 
the magnitudes of the momenta turn out to much larger than $B$; hence the wave 
function is much less affected by the presence of $B$. Indeed we find 
numerically that extended states are present if there is only an ``L"-shaped 
potential but no magnetic field, while a bound state can appear at the 
corner only if a magnetic field is applied.

{\bf Physical numbers:} We have presented all our numerical results in 
dimensionless units for convenience. However, we must convert these to some 
physical numbers in order to think of testing these results experimentally. 
To do this, let us fix the lattice spacing to be, say, $a= 0.1 ~\mu m$. In the
absence of a potential and a magnetic field, the dispersion of a massless 
Dirac electron with momentum $k_x$ is given, on a lattice, by $E=(\hbar v_F/a)
\sin (k_x a)$. For the topological insulator $\rm Bi_2 Se_3$, the velocity on 
the $x-y$ surface (perpendicular to the quintuple layers) is given 
by~\cite{tirev2} $\hbar v_F = 3.33 ~eV$\AA. This means that the values of
energy on the $y$-axis of Fig.~\ref{fig:disp} are in units of $\hbar 
v_F /a = 3.33 \times 10^{-3} ~eV$, and the values of $k_x$ on the $x$-axis of 
Figs.~\ref{fig:disp}, \ref{fig:decay} and \ref{fig:spin} are in units of $1/a 
= 10 ~\mu m^{-1}$. The decay length on the $y$-axis of Fig.~\ref{fig:decay} 
and the $x$ and $y$ coordinates in various figures are all
in units of $0.1 ~\mu m$. Next, a potential barrier of the form given in 
Eq.~\eqref{eq:barr} with $V_b = \pi/2$ and $\si = 2$ corresponds to a 
potential whose maximum value is $(V_b/\si \sqrt{2\pi}) (\hbar v_F/a) = 1.04 
\times 10^{-3} ~eV$ and width is $0.2 ~\mu m$. Finally, the Bohr magneton 
$\mu = e \hbar /(2m_e c) = 5.79 \times 10^{-5} ~ eV/T$. Assuming the 
gyromagnetic ratio to be $g=2$ as for a free electron, a value of $B = 
- g \mu B_z /2 = \pi/20$ corresponds to a magnetic field strength $|B_z| = 
(\pi/20) (\hbar v_F/a)/(5.79 \times 10^{-5}) = 9.03 ~T$. The numbers given 
above should only be considered to be rough guidelines; we expect our results 
for the bound states and their various properties to hold for a range of 
parameters.

\section{Summary and discussion}

In this paper, we have used a lattice model to study how a combination of 
time-reversal invariant (non-magnetic) potentials and a magnetic field can be 
used to confine Dirac electrons in different geometries. Our main results are 
as follows. 

\noi (i) For an infinitely long potential barrier and no magnetic field, the 
dispersion of the edge states propagating along the barrier is qualitatively 
of the form $E = v|k_x|$, where $k_x$ is the momentum along the barrier, 
if $|k_x|$ is much smaller than the inverse lattice spacing;
note that this is quite different from a chiral dispersion which is given 
by $E=vk_x$. The expectation value of the spin, $\la {\vec \si} \ra$, of the 
edge states lies in the $y$ direction. The velocity $v$ of the edge states
is smaller than that of the surface states and it can be varied by changing 
the strength of the potential barrier. The velocity $v$ becomes
very small for a particular value of the potential barrier, giving rise to an 
almost flat band near $E=0$. The wave function of the edge states decays 
exponentially away from the potential barrier; the decay length is inversely
proportional to $|k_x|$. Hence the edge states will cease to exist when 
the decay length becomes comparable to the size of the system.

\noi (ii) When a Zeeman field is applied in the $z$ direction, the surface 
states become gapped but the edge states do not. Further, an edge state now 
exists even for $k_x = 0$. The spin expectation value develops a component 
along the $z$ direction. Since the dispersion of the edge states can be 
controlled by the strength of the potential barrier, the edge states define 
a tunable one-dimensional system which is separated from the surface states 
by a gap. 

\noi (iii) Next we study what happens when there is a potential localized in 
some region. In the absence of a Zeeman field such a potential does not 
produce any localized states. But when a Zeeman field is turned on, we 
find that exponentially localized states can appear if the potential 
is strong enough. This gives us a zero-dimensional system. 

\noi (iv) We then study a combination of a long potential barrier, a localized
potential and a magnetic field; we find that states can appear which are bound
to the localized potential. We also study what happens if there is an 
``L"-shaped potential consisting of two semi-infinitely long arms meeting at 
a corner and a magnetic field. We find that there can be both states bound 
to the corner of the ``L" as well as scattering states along the arms.

Our results can be experimentally tested in a number of ways. To begin with, 
a potential barrier (straight or bent) can be produced by placing an
appropriately shaped gate close to the surface of
a TI and tuning the gate voltage. Then spin-resolved ARPES can be used to 
find the energy dispersion and spins of the different edge states. However, 
this method is not easy to use when a magnetic field is present
since the field would affect the trajectories of the electrons emitted
from the surface. A second method would be to measure the local density of
states using the tunneling conductance from a spin-polarized STM tip
which is placed very close to the barrier. If the local density of
states is found to be higher when a potential barrier is present compared 
to the case of no potential, this would provide evidence for the edge states.
An almost flat band would give rise to a particularly large density of
states at the location of the barrier.
Finally, it would be interesting to measure the differential conductance 
of the quasi-one-dimensional system which is produced by a long potential 
barrier (either straight or with bends as in an ``L"-shaped potential), and 
study how this varies with the potential barrier or a magnetic field; such
a variation would provide indirect evidence for the edge states. Note that 
since the edge states carry a spin (which is different for opposite edge 
momenta $+k_x$ and $-k_x$), a non-zero charge conductance along the barrier 
would also imply a non-zero spin conductance. 

We end by pointing out some directions for future work.
We have only considered the effects of a Zeeman coupling to a magnetic
field in this work. A magnetic field that has only a Zeeman coupling and no 
orbital coupling can be realized in a TI by doping with magnetic 
impurities~\cite{dopmag} or by depositing a ferromagnetic layer on the 
surface~\cite{feroti}. However, one should, in general, study the effects 
of the orbital coupling of electrons to a magnetic field. In a lattice model, 
such a coupling can be introduced through the phase in the couplings between 
nearest neighbor sites following the Peierls prescription.

Our work has shown that in the presence of a magnetic field, one can use 
potentials of various shapes to form wave guides along which Dirac electrons 
can propagate.
This idea may be used to construct a network of quantum wires by laying down 
appropriate potentials on the surface of a topological insulator. For this
purpose it would be useful to study the scattering matrix and conductance of
quasi-one-dimensional systems with ``L"-shaped bends and ``T"-junctions.

Finally, it would be interesting to study the effect of electron-electron 
interactions. The almost flat band of states that can be produced by tuning 
the barrier potential can be a platform for hosting a variety of strongly 
correlated electron states.
\vspace*{.6 cm}

\acknowledgments
We thank Oindrila Deb and Abhiram Soori for discussions. D.S. thanks DST, 
India for support under Grant No. SR/S2/JCB-44/2010.

\end{document}